\documentclass[10pt]{article}
\usepackage{amsmath,amssymb,amsthm,mathtools,bm}
\usepackage{graphicx,booktabs,enumitem,xcolor,array}
\usepackage{cite}
\usepackage{placeins}
\usepackage[colorlinks=true,linkcolor=blue!50!black,citecolor=blue!50!black,urlcolor=blue!50!black]{hyperref}
\usepackage{aliascnt}
\usepackage[capitalise,noabbrev]{cleveref}

\newtheorem{theorem}{Theorem}[section]
\newcommand{\sharedthm}[2]{\newaliascnt{#1}{theorem}\newtheorem{#1}[#1]{#2}\aliascntresetthe{#1}}
\sharedthm{proposition}{Proposition}
\sharedthm{lemma}{Lemma}
\sharedthm{corollary}{Corollary}
\theoremstyle{definition}
\sharedthm{definition}{Definition}
\theoremstyle{remark}
\sharedthm{remark}{Remark}
\sharedthm{example}{Example}
\numberwithin{equation}{section}

\makeatletter
\newcommand{\cnd}[2]{\hyperlink{cond-#1}{\textup{(#2)}}}
\newcommand{\cndef}[2]{\Hy@raisedlink{\hypertarget{cond-#1}{}}\textbf{(#2)}}
\makeatother
\IfFileExists{bbm.sty}{\usepackage{bbm}\newcommand{\ind}{\mathbbm{1}}}{\newcommand{\ind}{\mathrm{1\!\!1}}}
\newcommand{\R}{\mathbb{R}}
\newcommand{\E}{\mathsf{E}}
\newcommand{\Prob}{\mathsf{P}}
\newcommand{\tr}{\operatorname{tr}}
\newcommand{\G}{\mathcal{G}}
\newcommand{\Ms}{\mathcal{M}^{\mathrm{s}}_K}
\newcommand{\Sp}{\overline{\mathcal{S}}_K}
\newcommand{\cv}{\operatorname{cv}}
\newcommand{\Var}{\operatorname{\mathsf{Var}}}
\newcommand{\FN}{F_{\mathcal N}}

\begin{document}
\begin{center}
{\Large \textbf{Optimal Importance Sampling Densities for Estimating Several Expectations at Once}}

\vspace{0.5cm}
JASON BEH$^{1}$ \& J\'ER\^OME MORIO$^{2}$

{\footnotesize $^{1}$Computing \& Mathematical Sciences Division, Mohamed bin Zayed University of Artificial Intelligence, Abu Dhabi, UAE.}\\
{\footnotesize $^{2}$ONERA/DTIS, Universit\'e de Toulouse, FR.}\\
{\footnotesize E-Mail:\,}
\texttt{\emph{\footnotesize jason.kaixun@mbzuai.ac.ae, jerome.morio@onera.fr}}

\begin{abstract}
Simulation studies often estimate several quantities from a single sample, and the best importance sampling density depends on how the errors on these quantities are combined. We characterise the optimal auxiliary density for a general error criterion. We show that it is the nominal density reweighted by a weighted length of the vector of estimated functions, where the weights are set by the error criterion, or by the curvature of the loss when the criterion comes from a loss on the errors. We also provide a lower bound on the best achievable error. The optimal density is explicit for smooth losses, including the usual losses on covariance matrices. We also derive allocation laws for the cells of a histogram and characterise the minimax density for the probabilities of several events. When there is no closed form, we show how to compute the optimal density numerically.
\end{abstract}
\end{center}

\section{Introduction}\label{sec:intro}

A simulation study rarely estimates a single quantity. When the output of a model depends on random inputs $X\in\R^d$ with a known density $f$, one is usually interested in estimating several expectations under $f$. This arises, for instance, in reliability analysis, for several failure modes \cite{OMC19,BMSW25}, for the output distribution and its tails \cite{CL17}, or for the mean and covariance matrix that adaptive schemes re-estimate at each iteration \cite{UPMS21,BSS25}; in statistics, for a family of Bayes factors \cite{Dos10,RE24}; and in machine learning, for the gradients of a loss over the training examples \cite{ZZ15} or the probabilities of several rare outputs \cite{WH25}. A single sample is natural when each evaluation of the model is costly. Each of these problems concerns the estimation of a vector of expectations $m:=\E_f\{\psi(X)\}\in\R^K$, $\psi=(\psi_1,\dots,\psi_K)$.

Importance sampling (IS) draws $X_1,\dots,X_n$ from an auxiliary density $g$ and corrects for the change of law by the likelihood ratio $\ell:=f/g$. The resulting estimator $\widehat m_n:=n^{-1}\sum_i\ell(X_i)\psi(X_i)$ is unbiased, and its covariance matrix $C(g)/n$ depends on $g$. For a single quantity, $K=1$, the optimal choice of $g$ is known: $g\propto f|\psi|$ minimises the variance, which vanishes when $\psi\ge0$. For instance, to estimate a probability, $\psi=\ind_A$, this is $f|_A:=f\ind_A/\Prob_f(X\in A)$, the law of the inputs given $X\in A$. However, for a vector of expectations, one auxiliary density must serve all components, but no density is proportional to every $f|\psi_k|$ at once. The optimal choice of density then depends on how the errors on the components are combined.

We call error criterion a function $\Phi$ that aggregates the covariance matrix $C(g)$ into a scalar $\Phi\{C(g)\}$, to be minimised over $g$: for instance the sum of the variances $\tr C(g)$, the largest squared coefficient of variation $\max_kC_{kk}(g)/m_k^2$, or the asymptotic expectation of a loss $L(\widehat m_n-m)$ on the vector of errors, such as a norm or a divergence between histograms.

Practitioners combine several targets with mixtures \cite{OZ00,VG95,Owe13}, whose weights or weighting functions are optimal for a single quantity \cite{HO14,KVG19,MW96}. For one density serving a vector of expectations, optimal densities are known, to our knowledge, only for linear error criteria such as the sum of the variances, with optimum $f\|\psi\|$ \cite[Sec.~4.2.2]{LM25}, which is dominated by the largest components when their scales differ. The closest work is that of Demange-Chryst, Bachoc and Morio \cite{DBM23}, who combine adaptive importance sampling with control variates for the same problem. We instead characterise the optimal auxiliary density itself, for a general error criterion, and the trace optimum $f\|\psi\|$ is a special case. Our result could be combined with control variates. Buta and Doss \cite{BD11} and Roy and Evangelou \cite{RE24} choose proposal densities among candidates for a family of targets, by reducing the largest variance over the family or by criteria that are error criteria in our sense; we characterise the optimum over all densities.

\paragraph{Main result in brief.} This paper identifies the optimal auxiliary density for a given error criterion, and gives a lower bound against which any candidate density can be judged. For a convex error criterion $\Phi$, the optimal density is, whenever a minimiser and a matrix $M^\star$ as below exist, e.g.\ for coercive error criteria differentiable at the optimum, relative to $\Sp$, with a gradient $M^\star$ such that $M^\star\psi$ is not $f$-almost everywhere zero (\cref{prop:nec}),
\[
g^\star\ =\ \frac{f\sqrt{\psi^\top M^\star\psi}}{\E_f\big\{\sqrt{\psi(X)^\top M^\star\psi(X)}\big\}},\qquad M^\star\in\partial\Phi\big\{C(g^\star)\big\},
\]
the nominal density reweighted by the length of the vector of estimated functions in a metric $M^\star$, a (sub)gradient of the error criterion at the optimum, i.e.\ $M^\star=\nabla\Phi\{C(g^\star)\}$ when $\Phi$ is differentiable (\cref{thm:main}). For a loss $L$ on the vector of errors, $M^\star$ is half the Hessian of $L$ averaged over the Gaussian errors that $g^\star$ itself produces (\cref{thm:loss}). \Cref{tab:summary} gives them for the problems treated in \cref{sec:closed}.
\begin{table}[h]
\centering\small
\begin{tabular}{@{}lll@{}}
\toprule
quantities, error criterion & weight matrix $M^\star$ & optimal density\\
\midrule
any $\psi$, sum of the variances $\tr C$ & $I$ & $f\|\psi\|$ \cite{LM25}\\
any $\psi$, smooth loss with Hessian $H$ at $0$ & $H/2$ & $f\sqrt{\psi^\top H\psi}$\\
histogram, loss $\sum_k|x(k)|^r/p_k^\beta$ & diagonal & mass $\propto p_k^{2(r-\beta)/(r+2)}$ on rare cells\\
probabilities of events, $\max_kC_{kk}/m_k^2$ & $\mathrm{diag}(w^\star/m^2)$ & essentially the equal mixture of the $f|_{A_k}$\\
covariance of $f$, Stein loss & $\propto(\sigma_i\sigma_j)^{-1}$ & $f\;y^\top\Sigma^{-1}y$, $y=x-\E_f(X)$\\
probability and sensitivity, relative error & $\mathrm{diag}(1/m^2)/2$ & $f\ind_A\big\{m_1^{-2}+m_2^{-2}(\partial_\vartheta\log f)^2\big\}^{1/2}$\\
\bottomrule
\end{tabular}
\caption{Weight matrices and optimal densities for the problems treated in \cref{sec:closed}.}
\label{tab:summary}
\end{table}

\paragraph{Contributions.} We list our contributions as follows.
\begin{enumerate}[leftmargin=*]
\item \textbf{The optimal density and a lower bound.} We characterise the optimal density for a convex error criterion (\cref{thm:main}), and any guess $M$ of the weight matrix with $\Phi^*(M)<\infty$ yields a lower bound on the best achievable error, which can be estimated from a pilot sample. For a loss on the vector of errors, the weight matrix is the averaged curvature of the loss (\cref{thm:loss}), so that different losses lead to different densities; it can be computed from a pilot sample without new model evaluations (\cref{sec:comp}).
\item \textbf{Closed forms.} \Cref{tab:summary} summarises the closed forms. For a covariance matrix, the optimal density is explicit for the usual smooth losses (\cref{lem:local}); for a histogram, each cell should receive a share of the sample that is a power of its probability, from crude Monte Carlo for the squared error to equal shares for relative errors (\cref{prop:alloc}); and for the probabilities of several events in relative error, the equal mixture of their conditional densities is essentially optimal (\cref{thm:minimax,prop:ind}).
\item \textbf{What matters in practice.} The loss changes the optimal density by orders of magnitude for histograms, whereas for covariance matrices its influence fades as the dimension grows. In a reliability example, where the covariance matrix of the inputs given failure is estimated, as in adaptive schemes such as the cross-entropy method \cite{Rub97,RK04}, restricting the auxiliary density to Gaussian densities multiplies the quadratic risk by $2.2$ to $3.6$; the Gaussian fitted to this conditional law even has infinite risk, and mixtures of two Gaussian densities recover much of the gap (\cref{sec:closed,sec:num}).
\end{enumerate}

\paragraph{Organisation.} \Cref{sec:problem} states the problem and the conditions used throughout. \Cref{sec:convex} gives the optimal density for a convex error criterion and the dual bound, then shows that the problem is convex and has a solution. \Cref{sec:loss} gives the optimal density for a loss on the vector of errors, then justifies it. \Cref{sec:closed} gives closed forms with numerical illustrations, \cref{sec:num} computes the optimal densities for covariance matrices, where no closed form exists, together with the cost of parametric families, and \cref{sec:discussion} discusses when the error criterion matters and the consequences for adaptive schemes. Computational details are in \cref{app:num}, and all proofs are gathered in \cref{app:proofs}.

\section{The problem}\label{sec:problem}

This section states the optimisation problem over auxiliary densities and the conditions used throughout. The optimal density is given in \cref{sec:convex}, which also shows that the problem is convex and has a solution.

Throughout, the nominal density $f$ on $\R^d$ is arbitrary, and we use the following notation.
\begin{center}\small
\begin{tabular}{@{}ll@{}}
$x(1),x(2),\dots$ & coordinates of a point $x$; other vectors use subscripts, e.g.\ $m_k$\\
$p^2/\alpha$ & operations on vectors are coordinatewise\\
$\Ms$, $\mathcal S_K$, $\Sp$ & symmetric, positive-definite, positive-semidefinite $K\times K$ matrices\\
$C\preceq C'$ & Loewner order, i.e.\ $C'-C\in\Sp$\\
$\|M\|$, $\|M\|_F$ & operator and Frobenius norms of a matrix $M$\\
$e_k$, $\mathbf 1$, $\mathrm{diag}(v)$ & canonical basis vectors, vector of ones, diagonal matrix with diagonal $v$\\
$\Delta_K$ & simplex $\{w\in[0,1]^K:\sum_kw_k=1\}$\\
$\partial\Phi(C)$ & subdifferential $\{M\in\Ms:\Phi(C')\ge\Phi(C)+\tr\{M(C'-C)\}\text{ for any }C'\in\Ms\}$\\
$\lambda_{\min}(M)$, $\varphi_{a,\Gamma}$ & smallest eigenvalue of $M$, density of $\mathcal N(a,\Gamma)$
\end{tabular}
\end{center}\medskip

\label{sec:setting}

One sample from $g$ serves to estimate $K$ expectations at once; the quality of $g$ is summarised by the covariance matrix $C(g)$ of the estimator, and an error criterion $\Phi$ turns this matrix into a single number to be minimised over $g$. Let $\psi=(\psi_1,\dots,\psi_K):\R^d\to\R^K$ and $m:=\E_f\{\psi(X)\}$. With $X_1,\dots,X_n$ i.i.d.\ from $g$, the IS estimator $\widehat m_n:=n^{-1}\sum_{i\le n}\ell(X_i)\psi(X_i)$ has covariance $C(g)/n$, with
\begin{equation}\label{eq:C}
C(g):=S(g)-mm^\top\qquad\text{and}\qquad S(g):=\int\frac{f^2\,\psi\psi^\top}{g}.
\end{equation}
We denote by $\G$ the set of densities $g$ such that $g>0$ almost everywhere on $\{f\|\psi\|>0\}$ and $S(g)$ is finite, and we consider the problem
\begin{equation}\label{eq:problem}
\text{minimise }\ \Phi\big\{C(g)\big\}\ \text{ over }\ g\in\G .
\end{equation}
When $f$ is only known up to a constant, as for a Bayesian posterior, the self-normalised estimator $\sum_i\ell(X_i)\psi(X_i)/\sum_i\ell(X_i)$ has asymptotic covariance $\int f^2(\psi-m)(\psi-m)^\top/g$, which is $C(g)$ computed for the centred function $\psi-m$; everything below then applies with $(\psi,mm^\top)$ replaced by $(\psi-m,0)$.

Consider the following conditions:
\begin{itemize}[leftmargin=3.4em,labelindent=0pt,labelwidth=2.8em,labelsep=0.6em,align=left,itemsep=2pt]
\item[\cndef{I}{A1}] The functions $\psi$ are $f$-integrable, in the sense that
\begin{equation}\label{eq:I}
\int f\,\|\psi\|<\infty .
\end{equation}
\item[\cndef{C}{A2}] The error criterion $\Phi:\Ms\to(-\infty,\infty]$ is finite and nondecreasing for $\preceq$ on $\Sp$, and $\Phi=+\infty$ on $\Ms\setminus\Sp$.
\item[\cndef{Cplus}{A3}] Condition \cnd{C}{A2} is met and, in addition, $\Phi$ is convex and lower semicontinuous.
\end{itemize}
Condition \cnd{I}{A1} is the minimal requirement for $m$ to exist. Condition \cnd{C}{A2} says that a density with smaller covariance, in the Loewner order, is never worse. Condition \cnd{Cplus}{A3} is met by the trace $\tr C$, a weighted trace $\tr(WC)$ with $W\in\Sp$, the largest eigenvalue, and the largest squared coefficient of variation $\max_kC_{kk}/m_k^2$ when $m_k\neq0$ for any $k$.

\section{The optimal density for a convex error criterion}\label{sec:convex}

We now fix an error criterion $\Phi$ and look for the auxiliary density that minimises $\Phi\{C(g)\}$, that is, the best density when the errors on the components are aggregated through $\Phi$. We first introduce a family of candidate densities and a lower bound (\cref{sec:family}), and then state the optimality result (\cref{sec:main}).

\subsection{Weighted-length densities and the dual bound}\label{sec:family}

For a single function, $K=1$, the optimal density is $g\propto f|\psi|$. With several components no density can be proportional to every $f|\psi_k|$ at once, and the right replacement for $|\psi|$ turns out to be the length $\sqrt{\psi^\top M\psi}$ of the vector $\psi$ in a metric $M\in\Sp$ that weights the components against each other. Under condition \cnd{I}{A1}, the normalising constants
\begin{equation}\label{eq:gM}
Z(M):=\int f\sqrt{\psi^\top M\psi}\le\|M\|^{1/2}\int f\|\psi\|<\infty\qquad\text{and}\qquad g_M:=\frac{f\sqrt{\psi^\top M\psi}}{Z(M)}
\end{equation}
are well defined for any $M\in\Sp$ with $Z(M)>0$. The family $(g_M)$ does not depend on $\Phi$: the matrix $M$ is a set of priority weights, and $g_M$ puts mass where $\psi(x)$ is long in the metric $M$. For instance $M=I$ gives $g_I\propto f\|\psi\|$, and $M=e_ke_k^\top$ gives the optimal density $f|\psi_k|/\E_f\{|\psi_k(X)|\}$ for the $k$-th component alone. Only the direction of $M$ matters, since multiplying $M$ by a positive constant leaves $g_M$ unchanged. The error criterion enters only through the choice of $M$, made in \cref{sec:main}.

Each density $g_M$ is optimal for the linear error criterion $C\mapsto\tr(MC)$: by the Cauchy--Schwarz inequality, for any $g\in\G$,
\begin{equation}\label{eq:cs}
\tr\{MS(g)\}=\int\frac{f^2\,\psi^\top M\psi}{g}\ \ge\ \Big(\int f\sqrt{\psi^\top M\psi}\Big)^2\ =\ Z(M)^2,
\end{equation}
with equality if and only if $g=g_M$. A general convex error criterion is the supremum of the affine functions that lie below it, which leads to the following lower bound. We let $\Phi^*(M):=\sup_{C}\{\tr(MC)-\Phi(C)\}$ be the convex conjugate of $\Phi$, and we set
\begin{equation}\label{eq:Lambda}
\Lambda(M):=Z(M)^2-m^\top Mm-\Phi^*(M),\qquad M\in\Sp .
\end{equation}
We call $\Lambda(M)$ the dual bound: by the Fenchel--Young inequality and \eqref{eq:cs}, $\Phi\{C(g)\}\ge\tr\{MC(g)\}-\Phi^*(M)\ge\Lambda(M)$ for any $g$.

\subsection{Optimality}\label{sec:main}

The optimal density is the member of the family $(g_M)$ whose weight matrix is a gradient of the error criterion at its own covariance.

\begin{definition}\label{def:selfcons}
A matrix $M\in\Sp$ is self-consistent for the error criterion $\Phi$ if $Z(M)>0$, $g_M\in\G$ and
\[
M\ \in\ \partial\Phi\big\{C(g_M)\big\},
\]
that is, if $M$ is a (sub)gradient of $\Phi$ at the covariance of its own density $g_M$. We denote such a matrix by $M^\star$.
\end{definition}

Since $g_M$ is unchanged when $M$ is multiplied by a positive constant, the density $g_M$ only fixes the direction of $M$; self-consistency also fixes its scale. We are now in position to state the main result of the paper.

\begin{theorem}\label{thm:main}
Assume \cnd{I}{A1} and \cnd{Cplus}{A3}, and let $M^\star$ be self-consistent for $\Phi$. Then $g^\star:=g_{M^\star}$ is optimal and we have
\[
\min_{g\in\G}\Phi\{C(g)\}\ =\ \Phi\{C(g^\star)\}\ =\ \max_{M\in\Sp}\Lambda(M)\ =\ \Lambda(M^\star).
\]
In addition, without assuming that a self-consistent matrix exists, $\Lambda$ is concave on $\Sp$ and for any $g\in\G$ and $M\in\Sp$ we have
\begin{equation}\label{eq:weak}
\Phi\{C(g)\}\ \ge\ \Lambda(M).
\end{equation}
\end{theorem}

\Cref{thm:main} reduces an optimisation over densities to the choice of one $K\times K$ matrix, and turns any admissible guess of it, e.g.\ from a pilot sample, into a dual bound against which a practical density can be judged. Its structure parallels the Kiefer--Wolfowitz equivalence theorem of optimal experimental design \cite{KW60,Puk93}. There, a design measure is optimal when a gradient condition holds at its own information matrix; here, $g$ plays the role of the design measure and self-consistency that of the gradient condition. Unlike the information matrix, however, the covariance $C(g)$ is not linear in $g$.

The dual bound \eqref{eq:weak} is informative only when $\Phi^*(M)<\infty$, and two situations make $\Phi^*(M)$ explicit:
\begin{itemize}[leftmargin=*,itemsep=0pt]
\item If $\Phi$ is convex and differentiable at some $C'\in\mathcal S_K$ and $M:=\nabla\Phi(C')$, the Fenchel--Young inequality is an equality at $(C',M)$, so that $\Phi^*(M)=\tr(MC')-\Phi(C')$ and
\[
\Lambda(M)=Z(M)^2-m^\top Mm-\tr(MC')+\Phi(C').
\]
\item If $\Phi$ is positively homogeneous of degree one, as the four error criteria listed after condition \cnd{Cplus}{A3}, then $\Phi^*$ is $0$ on $\mathcal D_\Phi:=\{M:\tr(MC)\le\Phi(C)\text{ for any }C\in\Sp\}$ and $+\infty$ outside, so that $\Lambda(M)=Z(M)^2-m^\top Mm$ on $\mathcal D_\Phi$. For instance $\mathcal D_\Phi\cap\Sp=\{M\in\Sp:M\preceq W\}$ for $\tr(WC)$ and $\{M\in\Sp:\tr M\le1\}$ for the largest eigenvalue, while $\mathcal D_\Phi$ contains $\mathrm{diag}(w/m^2)$, $w\in\Delta_K$, for $\max_kC_{kk}/m_k^2$. A guess of $M$, e.g.\ estimated from a pilot sample, is then rescaled to the boundary of $\mathcal D_\Phi$.
\end{itemize}
In both cases $Z(M)$ and $m$ are expectations under $f$, so that $\Lambda(M)$ can be estimated from a pilot sample, for a matrix $M$ that may itself be estimated from this sample. The bound $\Lambda(M)\le\min_g\Phi\{C(g)\}$ is exact, but its pilot estimate is subject to Monte Carlo error: it estimates a lower bound rather than certifying one.

The following example shows the theorem at work in the simplest vector problem.

\begin{example}\label{ex:two}
Let $\psi_k:=\ind_{A_k}$, $k=1,2$, with $p_k:=\Prob_f(X\in A_k)>0$, and consider the weighted sum of squared coefficients of variation
\[
\Phi(C):=w_1\frac{C_{11}}{p_1^2}+w_2\frac{C_{22}}{p_2^2},\qquad w\in\Delta_2 .
\]
This error criterion is linear, with gradient $\mathrm{diag}(w_1/p_1^2,w_2/p_2^2)$ everywhere, so that $M^\star=\mathrm{diag}(w_1/p_1^2,w_2/p_2^2)$ and \cref{thm:main} yields
\[
g^\star\ \propto\ f\,\Big(w_1\frac{\ind_{A_1}}{p_1^2}+w_2\frac{\ind_{A_2}}{p_2^2}\Big)^{1/2}.
\]
When $A_1=A_2=A$, this is $f|_A$, the zero-variance density, and the optimal value $\Lambda(M^\star)$ is $0$. When $A_1$ and $A_2$ are disjoint, $g^\star=\alpha_1f|_{A_1}+\alpha_2f|_{A_2}$ with $\alpha_k\propto\sqrt{w_k}$, and the optimal value is $\Lambda(M^\star)=(\sqrt{w_1}+\sqrt{w_2})^2-1$; each event receives the same auxiliary mass no matter its probability when $w_1=w_2$. On an overlap $A_1\cap A_2$, the density $g^\star$ is proportional to $f\,(w_1/p_1^2+w_2/p_2^2)^{1/2}$, which is not of the form $f\,(\alpha_1/p_1+\alpha_2/p_2)$: the optimum is in general not a mixture of the conditional densities $f|_{A_k}$. \Cref{sec:minimax} quantifies this difference.
\end{example}

The form $g_M$ is also necessary, without convexity of $\Phi$. Since the optimal covariance may be singular, as for histograms, where the cell indicators sum to one, we use a notion of differentiability that allows boundary points of $\Sp$: we say that $\Phi$ is differentiable at $C\in\Sp$ relative to $\Sp$, with gradient $M\in\Ms$, if $\Phi(C')=\Phi(C)+\tr\{M(C'-C)\}+o(\|C'-C\|)$ as $C'\to C$ within $\Sp$. Since $\Sp-C$ spans $\Ms$, the gradient is unique, and it coincides with the usual gradient when $C\in\mathcal S_K$.

\begin{proposition}\label{prop:nec}
Assume \cnd{I}{A1} and \cnd{C}{A2}. Let $g^\star\in\G$ be a minimiser of $g\mapsto\Phi\{C(g)\}$ on $\G$, and assume that $\Phi$ is differentiable at $C(g^\star)$ relative to $\Sp$, with gradient $M$. Then $M\in\Sp$ and, if in addition $Z(M)>0$, we have
\[
g^\star\ =\ g_M\ =\ \frac{f\sqrt{\psi^\top M\psi}}{Z(M)},\qquad Z(M)=\int f\sqrt{\psi^\top M\psi} .
\]
\end{proposition}

The condition $Z(M)>0$ holds as soon as $M\psi$ is not $f$-almost everywhere zero, for instance when $M\neq0$ and $C(g^\star)\in\mathcal S_K$.

For the trace error criterion $\Phi(C)=\tr C$, the sum of the variances of the components, $M^\star=I$ and $g^\star\propto f\|\psi\|$ \cite{LM25}, which is $f|_A$ for $K=1$ and $\psi=\ind_A$. This error criterion weights the components according to their scale: when $m_k$ ranges over orders of magnitude, the smallest components receive little weight, which matters when their relative errors are of interest (\cref{sec:minimax,sec:sens}). With $f$ uniform over training examples and $\psi$ the vector of their gradients, $f\|\psi\|$ is the variance-minimising sampling of stochastic gradient methods \cite{ZZ15}.

\paragraph{Convexity and existence.} The problem \eqref{eq:problem} is convex as soon as the error criterion is, since the second-moment matrix $S(g)$ is convex in $g$: the error criterion of a mixture of two auxiliary densities is at most the average of their error criteria (\cref{lem:convex}). An optimal density exists as soon as the error criterion is coercive (\cref{prop:exist}); both results are stated and proved in \cref{app:proofs}.

\section{From a loss to its error criterion and optimal density}\label{sec:loss}

In applications the error criterion is rarely given. One rather specifies a loss $L(\widehat m_n-m)$ on the vector of errors, such as a norm, a divergence between histograms or a loss on covariance matrices. Throughout, the argument $x\in\R^K$ of $L$ stands for the vector of errors $\widehat m_n-m$, so that $x(k)=\widehat m_{n,k}-m_k$ is the error on the $k$-th component. We show that every such loss defines an error criterion $\Phi_L$, to which \cref{sec:convex} applies. Consider the following condition on a loss function $L:\R^K\to[0,\infty)$:
\begin{itemize}[leftmargin=3.4em,labelindent=0pt,labelwidth=2.8em,labelsep=0.6em,align=left,itemsep=2pt]
\item[\cndef{L}{B1}] The loss $L$ is even and quasi-convex, $L(0)=0$, and one of the following holds:
\begin{enumerate}[label=(\alph*),leftmargin=1.6em,itemsep=0pt]
\item The loss $L$ is continuous and positively homogeneous of degree $\kappa>0$; we then set $L_0:=L$.
\item The loss $L$ is twice differentiable at $0$ with Hessian $H$, and $L(x)\le c_0\,(1+\|x\|)^{c_0}$ for some $c_0>0$; we then set $\kappa:=2$ and $L_0(x):=x^\top Hx/2$.
\end{enumerate}
\end{itemize}
Condition \cnd{L}{B1} covers norms and their powers (case (a)) and smooth even losses, such as weighted squared errors (case (b)); in case (b), $H\in\Sp$ because $L$ is minimal at $0$.

\subsection{The error criterion of a loss and its optimal density}\label{sec:PhiL}\label{sec:lossopt}

The link between a loss and an error criterion is the central limit theorem: $\sqrt n(\widehat m_n-m)$ is approximately $\mathcal N\{0,C(g)\}$, so that the error $\widehat m_n-m$ is of order $n^{-1/2}$, the loss $L(\widehat m_n-m)$ is of order $n^{-\kappa/2}$, and its expectation is asymptotically that of a Gaussian vector with covariance $C(g)$. For $C\in\Sp$ we therefore set
\begin{equation}\label{eq:PhiL}
\Phi_L(C):=\E\big\{L_0(C^{1/2}\xi)\big\},\qquad \xi\sim\mathcal N(0,I_K),
\end{equation}
and we call $\Phi_L\{C(g)\}$ the risk of the auxiliary density $g$ for the loss $L$; \cref{lem:risk} makes this precise. The Gaussian law in \eqref{eq:PhiL} is the limiting law of the errors given by the central limit theorem, for any nominal density $f$: no result of this paper assumes $f$ Gaussian. Applying \cref{sec:convex} to $\Phi=\Phi_L$ gives the optimal density for a loss.

\begin{theorem}\label{thm:loss}
Assume \cnd{I}{A1} and \cnd{L}{B1}. Let $g^\star_L\in\G$ be a minimiser of $g\mapsto\Phi_L\{C(g)\}$ on $\G$ such that $\Phi_L$ is differentiable at $C(g^\star_L)$ relative to $\Sp$, with gradient $M^\star_L$. Then $M^\star_L\in\Sp$ and, whenever $Z(M^\star_L)>0$, we have
\[
g^\star_L\ =\ \frac{f\sqrt{\psi^\top M^\star_L\psi}}{\E_f\big\{\sqrt{\psi(X)^\top M^\star_L\psi(X)}\big\}} .
\]
When $C(g^\star_L)\in\mathcal S_K$, the gradient is $M^\star_L=\E\{\nabla^2L_0(X^\star_L)\}/2$ with $X^\star_L\sim\mathcal N\{0,C(g^\star_L)\}$, where $\nabla^2L_0$ is the Hessian of $L_0$ in the sense of distributions. In addition, if $\Phi_L$ is convex on $\Sp$, then any $g_M\in\G$ such that $\Phi_L$ is differentiable at $C(g_M)$ relative to $\Sp$ with gradient $M$ is a global minimiser, and $\Phi_L\{C(g_M)\}=\Lambda(M)$.
\end{theorem}

A minimiser exists as soon as $f\|\psi\|$ is not almost everywhere zero and $L>0$ on $\R^K\setminus\{0\}$ in case (a), or $H\in\mathcal S_K$ in case (b) (\cref{prop:exist,lem:price}). Differentiability relative to $\Sp$ holds at any $C\in\mathcal S_K$, and on the boundary of $\Sp$ for quadratic losses and for sums $L_0(x)=\sum_jc_j|v_j^\top x|^r$ at any $C$ with $v_j^\top Cv_j>0$ for any $j$ (\cref{lem:price}).

\Cref{thm:loss} says that the optimal density $g^\star_L$ for a loss $L$ weights $f$ by $\sqrt{\psi^\top M^\star_L\psi}$, where $M^\star_L$ is half the Hessian of the loss averaged over the Gaussian error law that this density itself produces. It is a fixed point in one $K\times K$ matrix, whose size depends on the number $K$ of estimated quantities rather than on the dimension $d$ of the inputs, although $K=d(d+1)/2$ for covariance matrices; when $\Phi_L$ is convex, $M^\star_L$ is the self-consistent matrix of \cref{def:selfcons} for the error criterion $\Phi=\Phi_L$. This matrix, and hence the optimal density, depends on the loss: for the same vector $\psi$, two losses with different curvatures lead to different auxiliary densities. \Cref{sec:closed,sec:num} quantify this dependence.

For a quadratic loss the curvature is constant and $M^\star_L=H/2$. For $L(x)=\sum_k|x(k)|^r$, the matrix $M^\star_L$ is diagonal with $(M^\star_L)_{kk}\propto C_{kk}^{(r-2)/2}$. When $r<2$ the loss is steepest near zero, so that the components that are already estimated accurately receive more weight; when $r>2$ the weight goes to the components with the largest errors. This mechanism produces the allocation law of \cref{prop:alloc}.

The error criterion $\Phi_L$ is convex for quadratic losses, as are the error criteria listed after condition \cnd{Cplus}{A3}, but convexity can fail for norms: for $L(x)=\sum_k|x(k)|$, $\Phi_L(C)\propto\sum_k\sqrt{C_{kk}}$ is concave. The first part of \cref{thm:loss} is then a necessary condition only. Wherever a brute-force minimisation was feasible, for the partitions of \cref{sec:alloc}, it returned the fixed point; for the operator and nuclear norms of \cref{sec:num} we could not certify global optimality.

\paragraph{Why the theorem holds.} \Cref{thm:loss} combines \cref{thm:main,prop:nec} with three properties of $\Phi_L$, stated and proved in \cref{app:proofs}. First, $\Phi_L$ is the limit of the rescaled risk, $n^{\kappa/2}\,\E\{L(\widehat m_n-m)\}\to\Phi_L\{C(g)\}$, under a uniform integrability condition (\cref{lem:risk}). Second, by Anderson's inequality and Price's theorem \cite{And55,Pri58,Ste81}, $\Phi_L$ is nondecreasing for $\preceq$, and its gradient is half the averaged Hessian of the loss (\cref{lem:price}): for any $C\in\mathcal S_K$,
\begin{equation}\label{eq:price}
\nabla\Phi_L(C)\ =\ \tfrac12\,\E\big\{\nabla^2L_0(X)\big\}\ =\ \tfrac12\,C^{-1}\,\E\big\{L_0(X)\,(XX^\top-C)\big\}\,C^{-1}\ \in\ \Sp,\qquad X\sim\mathcal N(0,C),
\end{equation}
where $\nabla^2L_0$ stands for the Hessian of $L_0$ in the sense of distributions. Third, for the densities $g_M$ the weighted functions are bounded, $\|\ell\,\psi\|\le Z(M)\,\lambda_{\min}(M)^{-1/2}$, so that the limit of the risk holds whenever $M$ is positive definite (\cref{prop:ui}).

\subsection{Computation}\label{sec:comp}

The fixed point of \cref{thm:loss} can be computed from a single pilot sample, without new evaluations of the model. We iterate $M\leftarrow\nabla\Phi_L\{C(g_M)\}$ with damping; we do not prove that this iteration converges, but it did in all our examples. The gradient \eqref{eq:price} is available
\begin{itemize}[leftmargin=*,itemsep=0pt]
\item in closed form for quadratic and separable losses: $L(x)=\sum_k|x(k)|^r$ gives the diagonal matrix $M_{kk}=(r/2)\,\E(|\zeta|^r)\,C_{kk}^{(r-2)/2}$, with $\zeta\sim\mathcal N(0,1)$; for instance $M_{kk}=(2\pi C_{kk})^{-1/2}$ when $r=1$;
\item exactly for sums of absolute linear forms: $L(x)=\sum_j|v_j^\top x|$, with $v_j\in\R^K$, gives $M=\sum_jv_jv_j^\top/(2\pi\,v_j^\top Cv_j)^{1/2}$;
\item for maxima such as the operator norm, through the Hessian of a smooth approximation of the maximum (a log-sum-exp), averaged over Gaussian draws;
\item otherwise, through the last expression in \eqref{eq:price}, which only requires evaluations of $L_0$ at Gaussian draws (unbiased, but noisy when $C$ is ill-conditioned).
\end{itemize}
In practice $C(g_M)$ need not be computed by sampling $g_M$: since
\begin{equation}\label{eq:CgM}
C(g_M)=Z(M)\int\frac{f\,\psi\psi^\top}{\sqrt{\psi^\top M\psi}}-mm^\top,
\end{equation}
the integrals can be estimated by IS from a single pilot sample. The fixed-point iteration is therefore free in terms of model evaluations: it reuses the values of $\psi$ already computed on the pilot sample, and its cost is that of weighted averages over this sample and of operations on $K\times K$ matrices.

\begin{remark}[Use with a pilot sample]\label{rem:pilot}
Learning $M$ from a pilot sample is cheap, and sampling $g_M$ is the bottleneck. In practice one draws a pilot sample from a density $g_0$ that charges every region where $f\psi$ does, for instance an equal mixture of $f$ and of the intermediate densities of an adaptive scheme; estimates $m$, then $M$, which equals $H/2$ for a smooth loss and is otherwise obtained by the damped iteration $M\leftarrow[M+\nabla\Phi_L\{C(g_M)\}]/2$ from $M=I$; and draws the final sample from an approximation of $g_M$: exactly, by rejection from a mixture, when $\sqrt{\psi^\top M\psi}$ is piecewise constant, by Markov chain Monte Carlo, or from a parametric density mixed with $f$ to keep the likelihood ratio bounded. With a parametric sampler, only the pilot and the final sample evaluate the model, whereas rejection and Markov chain Monte Carlo evaluate it at every proposed point. The dual bound $\Lambda$ at $M':=\nabla\Phi_L(\widehat C)$, with $\widehat C$ the last estimate of $C(g_M)$, estimates from the pilot sample the most that a better density could gain. On two histograms of $u^\top X$ with $X\sim\mathcal N(0,I_{10})$ (\cref{app:pilot}), the weight matrix learned from $2000$ pilot draws gave a regret (risk divided by the minimal risk, as in \cref{sec:closed}) of at most $1.0015$, and, for the $\chi^2$ loss, the estimated dual bound had a relative spread of $3\%$ (the total variation loss has a concave $\Phi_L$ and no informative dual bound). The bottleneck is the sampler: with one Gaussian density per cell, the final risk exceeded that of $g_M$ by a factor $1.3$ to $1.9$. At equal total budget, including the evaluations spent on the adaptive scheme (here cross-entropy) and on the pilot sample, the resulting estimator reduced the risk of the equal mixture of the same Gaussian components by a factor $1.4$ to $2.2$ and that of the cross-entropy density by a factor $2$ to $19$, but beat crude Monte Carlo only for the $\chi^2$ loss at the largest budget, by a factor $1.35$.
\end{remark}

\Cref{sec:closed} makes \cref{thm:loss} explicit in three situations of practical interest, and \cref{sec:num} computes the optimal densities where no closed form exists.

\section{Closed-form expressions}\label{sec:closed}

\Cref{thm:loss} characterises the optimal density through a fixed point in the matrix $M$. This section identifies three situations where this fixed point is explicit or nearly so: smooth losses, where $M^\star_L$ does not depend on $g$ (\cref{sec:smooth}); partitions with separable losses, where the optimal density is a mixture of conditional densities (\cref{sec:alloc}); and probabilities of events in relative error, where the minimax density solves a concave problem on the simplex (\cref{sec:minimax}). It ends with a worked example from reliability analysis: a probability and its sensitivity to a distribution parameter (\cref{sec:sens}).

All numerical comparisons in this section and the next are ratios of risks. Since the risk of IS with $n$ draws is approximately $\Phi_L\{C(g)\}\,n^{-\kappa/2}$, a density with $\rho$ times the risk of another needs $\rho^{2/\kappa}$ times as many draws to reach the same accuracy. For the quadratic losses and the squared coefficients of variation used in these sections, $\kappa=2$ and $\rho$ is directly the factor on the sample size, that is, on the number of evaluations of the model; for the operator and nuclear norms, $\kappa=1$ and the factor is $\rho^2$. We call regret of a density, for a given loss, the ratio of its risk to the minimal risk for that loss. Where the candidate family is finite-dimensional, fixed points of \cref{thm:loss} were compared with a brute-force minimisation over the whole family; they coincided in every case. All examples take $f=\mathcal N(0,I)$, the usual setting in reliability analysis after an isoprobabilistic transformation, i.e.\ a change of variables that maps the inputs to standard Gaussian ones; computational details and Monte Carlo checks are in \cref{app:num}.

\subsection{Smooth losses}\label{sec:smooth}

We start with losses that are twice differentiable at $0$, which include squared errors and smooth divergences between estimate and truth. For the latter, which need not be even and then do not satisfy condition \cnd{L}{B1}, we take as error criterion $\Phi_L(C)=\tr(HC)/2$, the risk under the limiting Gaussian law of the errors. If $L$ is smooth at $0$, with Hessian $H$ at $0$ as in case (b) of condition \cnd{L}{B1}, then $\Phi_L(C)=\tr(HC)/2$ and $M^\star_L=H/2$ does not depend on $g$: the optimal density is explicit, $g^\star\propto f\sqrt{\psi^\top H\psi}$, as soon as $H\in\mathcal S_K$ and $f\|\psi\|$ is not almost everywhere zero (\cref{prop:ui}); for a singular $H$, $g_H$ may fail to be admissible and the infimum $\Lambda(H/2)$ need not be attained.

\paragraph{Covariance matrices.} We consider the estimation of the covariance matrix $\Sigma:=\Var_f(X)$ of the nominal density $f$, assumed to have finite second moments, mean $\mu$ and positive-definite $\Sigma$, by IS from an auxiliary density $g$; $f$ may be a posterior, the law of the inputs given failure, or the current target of an adaptive scheme. We take $\mu$ and the normalising constant of $f$ known, so that the likelihood ratio $f/g$ is available; otherwise, the self-normalised version of \cref{sec:setting} applies, with $\psi$ replaced by its centred version.

Since $\Sigma$ is an expectation under $f$, $\psi(x)$ is the vector of the $K=d(d+1)/2$ entries $i\le j$ of $(x-\mu)(x-\mu)^\top$, the off-diagonal ones multiplied by $\sqrt2$. We work in the eigenbasis of $\Sigma$, so that $\Sigma=\mathrm{diag}(\sigma_1,\dots,\sigma_d)$, where the eigenvalues $\sigma_i$ are the variances of $f$ along its principal axes; we write $y:=x-\mu$ in this basis, and $\delta:=\widehat\Sigma-\Sigma$ for the estimation error. For $d=2$, for instance, $K=3$ and $\psi(x)=(y(1)^2,\sqrt2\,y(1)y(2),y(2)^2)$; the factor $\sqrt2$ makes $\|\psi(x)\|$ equal to the Frobenius norm of $yy^\top$. In this encoding the error vector $\widehat m_n-m\in\R^K$ has coordinates $\delta_{ii}$ and $\sqrt2\,\delta_{ij}$, $i<j$, so that its Euclidean norm is $\|\delta\|_F$ and $\sum_{i,j}h_{ij}\delta_{ij}^2=\sum_{i\le j}h_{ij}(\widehat m_n-m)_{ij}^2$ for any symmetric array $(h_{ij})$. Covariance losses are usually written as functions of the estimate $\widehat\Sigma$, and we use those of Donoho, Gavish and Johnstone \cite{DGJ18}, in their fixed-dimension limit; through $\delta=\widehat\Sigma-\Sigma$, each of them defines a loss $L$ on the error vector, with $L(0)=0$. Their local forms are as follows.

\begin{lemma}\label{lem:local}
For each loss in \cref{tab:covloss}, seen as a function $L$ of $\delta$, we have
\[
L(\delta)\ =\ \sum_{i,j\le d}h_{ij}\,\delta_{ij}^2+o\big(\|\delta\|^2\big)\qquad\text{as }\delta\to0,
\]
with $(h_{ij})$ given in the second column of the table. Consequently $M^\star_L=\mathrm{diag}\{(h_{ij})_{i\le j}\}$ and\[\psi^\top M^\star_L\psi=\sum_{i,j}h_{ij}\,y(i)^2y(j)^2,\]which yields the optimal densities of the third column.
\end{lemma}

\begin{table}[h]
\centering\small
\begin{tabular}{@{}>{\raggedright\arraybackslash}p{5.4cm}ll@{}}
\toprule
loss & $h_{ij}$ & $g^\star\propto f\,\times$\\
\midrule
Frobenius $\|\delta\|_F^2$ & $1$ & $\|y\|^2$\\
Stein $\tr(\Sigma^{-1}\widehat\Sigma)-\log\det(\Sigma^{-1}\widehat\Sigma)-d$ & $(2\sigma_i\sigma_j)^{-1}$ & $y^\top\Sigma^{-1}y$\\
Kullback--Leibler $\mathrm{KL}\{\mathcal N(\mu,\widehat\Sigma)\|\mathcal N(\mu,\Sigma)\}$ & $(4\sigma_i\sigma_j)^{-1}$ & $y^\top\Sigma^{-1}y$\\
Bhattacharyya distance between the same laws & $(16\sigma_i\sigma_j)^{-1}$ & $y^\top\Sigma^{-1}y$\\
Fr\'echet (squared Wasserstein-2 between the same laws) & $\{2(\sigma_i+\sigma_j)\}^{-1}$ & $\big\{\sum_{i,j}y(i)^2y(j)^2/(\sigma_i+\sigma_j)\big\}^{1/2}$\\
precision $\|\widehat\Sigma^{-1}-\Sigma^{-1}\|_F^2$ & $(\sigma_i\sigma_j)^{-2}$ & $y^\top\Sigma^{-2}y$\\
log-Euclidean $\|\log\widehat\Sigma-\log\Sigma\|_F^2$ & $\{(\log\sigma_i-\log\sigma_j)/(\sigma_i-\sigma_j)\}^2$\,${}^{(*)}$ & $\big\{\sum_{i,j}h_{ij}y(i)^2y(j)^2\big\}^{1/2}$\\
\bottomrule
\end{tabular}
{\small ${}^{(*)}$With the convention $\sigma_i^{-2}$ when $\sigma_i=\sigma_j$.}
\caption{Local forms $h_{ij}$ of the covariance losses and optimal densities of \cref{lem:local}.}
\label{tab:covloss}
\end{table}

\noindent Except for the Frobenius loss, these losses are infinite or undefined when $\widehat\Sigma$ is not positive definite, which an IS estimate need not be, and they are not even, so that they do not satisfy condition \cnd{L}{B1} globally and \cref{lem:risk} does not apply to them. As stated above, their error criterion is $\Phi_L(C)=\tr(HC)/2$, with $H=2\,\mathrm{diag}\{(h_{ij})\}$, the usual definition of the asymptotic risk of such losses; the finite-sample risk itself may be infinite, e.g.\ for the precision loss when $\widehat\Sigma^{-1}$ is not integrable. \Cref{thm:main} applies directly to this linear error criterion. The Stein, Kullback--Leibler and Bhattacharyya losses differ locally by a constant factor and lead to the same density. The losses measure different things. The Frobenius loss measures entrywise errors, no matter the scale of $\Sigma$. The Stein, Kullback--Leibler and Bhattacharyya losses measure errors relative to $\Sigma$, and are natural when $\widehat\Sigma$ is used to define a Gaussian density, as in adaptive schemes. The Fr\'echet loss is the squared Wasserstein-2 distance between the corresponding Gaussian laws, and the log-Euclidean loss compares the matrix logarithms, which treats over- and under-estimation of an eigenvalue symmetrically. The precision loss matters when $\Sigma^{-1}$ is used, e.g.\ in a Gaussian log-density or a Mahalanobis distance; it is dominated by the small eigenvalues of $\Sigma$. The operator and nuclear norms are not smooth; they are treated in \cref{sec:num}.

\subsection{An allocation law for rare cells}\label{sec:alloc}

We now estimate the probabilities of the cells of a partition, as for a histogram, with a loss that is separable across cells. Let $(A_k)_{k\le K}$ be a partition of $\R^d$ into cells with probabilities $p_k:=\Prob_f(X\in A_k)>0$, and let $\psi_k:=\ind_{A_k}$, so that $\widehat m_n$ estimates the vector $p$ of cell probabilities, e.g.\ a histogram of the output of a model. We call a cell rare when $p_k\ll1$: the tail bins of a histogram, or the classes of extreme outputs. Estimating their probabilities together with those of the central cells is the vector analogue of rare event estimation, and the question is how much auxiliary mass each cell should receive. For $\alpha\in\Delta_K$, let $g_\alpha:=\sum_k\alpha_k\,f|_{A_k}$ be the mixture that gives mass $\alpha_k$ to cell $k$; crude Monte Carlo corresponds to $\alpha=p$. For these mixtures we have $C(g_\alpha)=\mathrm{diag}(p^2/\alpha)-pp^\top$, so that the relative error on cell $k$, measured by the coefficient of variation $\cv(Y):=\Var(Y)^{1/2}/|\E(Y)|$, is
\begin{equation}\label{eq:cell-cv}
n\,\cv(\widehat m_{n,k})^2\ =\ \frac{1-\alpha_k}{\alpha_k}.
\end{equation}
Smooth losses are covered by \cref{sec:smooth}: the KL divergence (either direction, in its generalised form $\sum_k\{\widehat m_{n,k}\log(\widehat m_{n,k}/p_k)-\widehat m_{n,k}+p_k\}$ and its reverse, since the IS estimates need not sum to one), $\chi^2$ and squared Hellinger distance between $\widehat m_n$ and $p$ all have local Hessian $H\propto\mathrm{diag}(1/p)$, and since $\psi^\top H\psi\propto1/p_k$ on $A_k$, the optimal density $f\sqrt{\psi^\top H\psi}$ gives cell $k$ auxiliary mass $\alpha_k\propto\sqrt{p_k}$. This square-root allocation lies between crude Monte Carlo, $\alpha=p$, and equal shares. The next result shows that it is one point of a one-parameter family.

\begin{proposition}\label{prop:alloc}
Let $\psi=(\ind_{A_1},\dots,\ind_{A_K})$ for a partition $A_1,\dots,A_K$ of $\R^d$ with $K\ge2$ and $p_k>0$, and assume $L(x)=\sum_k|x(k)|^r/p_k^\beta$, where $x(k)=\widehat m_{n,k}-p_k$ is the error on cell $k$, for some $r\ge1$ and $\beta\in\R$, so that condition \cnd{L}{B1} is met. Then any density $g^\star$ satisfying the conclusion of \cref{thm:loss} is a mixture $g^\star=g_\alpha$, and its weights satisfy, for any $k\le K$,
\begin{equation}\label{eq:alloc-exact}
\alpha_k^{(r+2)/4}\,(1-\alpha_k)^{-(r-2)/4}\ \propto\ p_k^{(r-\beta)/2}.
\end{equation}
In particular, on cells with $\alpha_k\ll1$ we have
\[
\alpha_k\ \propto\ p_k^{\,\eta(r,\beta)}\qquad\text{with}\qquad \eta(r,\beta):=\frac{2(r-\beta)}{r+2}.
\]
\end{proposition}
The approximation $\alpha_k\propto p_k^{\eta(r,\beta)}$ applies to cells with small auxiliary mass, that is, to rare cells when $\eta(r,\beta)>0$, and to every cell when $K$ is large and no single cell carries most of the mass. By \eqref{eq:cell-cv}, the squared relative error on a rare cell is then of order $p_k^{-\eta(r,\beta)}$: the exponent measures how fast the accuracy of the optimal density is allowed to degrade as cells become rarer. Crude Monte Carlo has $\eta=1$, so that $n\cv(\widehat m_{n,k})^2\approx1/p_k$; a loss with $\eta<1$ gives rare cells more mass than $f$, and a loss with $\eta=0$ gives every cell the same relative accuracy (\cref{tab:alloc}).
\begin{table}[h]
\centering\small
\begin{tabular}{@{}lccl@{}}
\toprule
loss & $(r,\beta)$ & $\eta$ & rare cells receive\\
\midrule
squared error & $(2,0)$ & $1$ & as under $f$: crude Monte Carlo is optimal\\
total variation & $(1,0)$ & $2/3$ & more than under $f$\\
KL, $\chi^2$, Hellinger & $(2,1)$ & $1/2$ & $\propto\sqrt{p_k}$\\
any relative-error loss $\sum_k|x(k)/p_k|^r$ & $(r,r)$ & $0$ & equal mass\\
fourth-power error & $(4,0)$ & $4/3$ & less than under $f$\\
\bottomrule
\end{tabular}
\caption{Exponent $\eta=\eta(r,\beta)$ of the allocation law of \cref{prop:alloc} for usual losses on histograms.}
\label{tab:alloc}
\end{table}

For a partition, every density is dominated by the mixture $g_\alpha$ with the same cell masses $\alpha_k=\int_{A_k}g$: the matrix $S(g)$ is diagonal and, by the Cauchy--Schwarz inequality, $S(g)_{kk}=\int_{A_k}f^2/g\ge p_k^2/\alpha_k$, so that $C(g)\succeq C(g_\alpha)$. Minimising over $\alpha\in\Delta_K$ is therefore a minimisation over all densities, which also covers the max-type losses below, to which \cref{thm:loss} does not apply directly; the maximum relative error is moreover covered by \cref{thm:minimax}. Note also that $C(g_\alpha)$ is singular, with kernel spanned by $\alpha/p$, which is why \cref{thm:loss} is stated with differentiability relative to $\Sp$.

\paragraph{Histograms.} As an illustration of \cref{prop:alloc}, for the cells $(-\infty,0),[0,1),\dots,[4,5),[5,\infty)$ of $\mathcal N(0,1)$, \cref{tab:hist} lists the optimal masses. The rare-cell ratios follow \cref{prop:alloc} (total variation: ratio of the masses of $[4,5)$ and $[3,4)$ equal to $0.083$ observed and predicted). The relative-error losses are $\sum_k|x(k)/p_k|^2$ and $\max_k|x(k)/p_k|$. Losses on the cumulative distribution function, whose error at cell $k$ is $\sum_{j\le k}x(j)$, namely the cumulative $L^1$ loss $\sum_k|\sum_{j\le k}x(j)|$ and the Kolmogorov loss $\max_k|\sum_{j\le k}x(j)|$, are dominated by the central cells, and the Kolmogorov loss even gives the tails less mass than $f$. The loss changes the answer by orders of magnitude: the squared-error optimum (crude Monte Carlo) is $8\times10^4$ times worse than optimal for the squared relative-error loss. For the $\chi^2$, total variation and cumulative $L^1$ losses, whose risks are explicit functions of $\alpha$, a minimisation from $200$ random starting points (respectively $50$) reproduced the optima of \cref{tab:hist} to machine precision, although the error criteria of the last two are not convex; for the Kolmogorov loss, whose risk is estimated by Monte Carlo, the fixed point was confirmed by a local minimisation over $\alpha$ only.

\begin{table}[h]
\centering\small
\begin{tabular}{@{}lccccccc@{}}
\toprule
loss & $(-\infty,0)$ & $[0,1)$ & $[1,2)$ & $[2,3)$ & $[3,4)$ & $[4,5)$ & $[5,\infty)$\\
\midrule
$f$, i.e.\ $\alpha=p$ & 0.500 & 0.341 & 0.136 & 0.021 & 0.0013 & $3.1\times10^{-5}$ & $2.9\times10^{-7}$\\
squared error & 0.500 & 0.341 & 0.136 & 0.021 & 0.0013 & $3.1\times10^{-5}$ & $2.9\times10^{-7}$\\
KL / $\chi^2$ / Hellinger & 0.382 & 0.316 & 0.199 & 0.079 & 0.020 & 0.0030 & $2.9\times10^{-4}$\\
total variation & 0.453 & 0.328 & 0.165 & 0.046 & 0.0071 & $5.9\times10^{-4}$ & $2.6\times10^{-5}$\\
relative (squared or max) & 0.143 & 0.143 & 0.143 & 0.143 & 0.143 & 0.143 & 0.143\\
Kolmogorov & 0.598 & 0.307 & 0.084 & 0.010 & $5.5\times10^{-4}$ & $1.4\times10^{-5}$ & $1.1\times10^{-7}$\\
\bottomrule
\end{tabular}
\caption{Optimal auxiliary mass per cell for a histogram of $\mathcal N(0,1)$. The cumulative $L^1$ loss gives essentially crude Monte Carlo (risk $0.1\%$ below it).}
\label{tab:hist}
\end{table}

Relative-error losses ($\beta=r$), which measure the error on each cell relative to its probability, all lead to equal mass. Their $r\to\infty$ limit is the minimax relative error, studied next for sets that may overlap.

\subsection{Probabilities of several events in relative error}\label{sec:minimax}

We next consider the probabilities of $K$ events $A_k$, such as failure modes, with sets that may overlap, and the largest squared coefficient of variation $\max_kC_{kk}/m_k^2$ as error criterion. It is the limit $r\to\infty$ of the error criteria of the relative-error losses $\sum_k|x(k)/m_k|^r$ of \cref{sec:alloc} after normalisation, since $\{\sum_k(C_{kk}/m_k^2)^{r/2}\}^{2/r}\to\max_kC_{kk}/m_k^2$, and it does not require the sets to form a partition. Recall that $\psi_k$ is the $k$-th estimated function and $m_k=\E_f\{\psi_k(X)\}$ its expectation; for probabilities, $\psi_k=\ind_{A_k}$ and $m_k=\Prob_f(X\in A_k)$. Assume $m_k\neq0$ for any $k\le K$. We set $V_k(g):=C_{kk}(g)/m_k^2=n\cv(\widehat m_{n,k})^2$, and for any $w\in\Delta_K$
\begin{equation}\label{eq:Theta}
Q_w:=\sum_{k\le K}w_k\,\frac{\psi_k^2}{m_k^2}\qquad\text{and}\qquad\Theta(w):=\E_f\big\{\sqrt{Q_w(X)}\big\}.
\end{equation}
The following result reduces the minimax problem to the maximisation of the concave function $\Theta$ over the simplex.

\begin{theorem}\label{thm:minimax}
Assume \cnd{I}{A1}, and $m_k\neq0$ for any $k\le K$. Then $\Theta$ is concave on $\Delta_K$. In addition, for any maximiser $w^\star$ of $\Theta$ on $\Delta_K$, the density $g^\star:=f\sqrt{Q_{w^\star}}/\Theta(w^\star)$ belongs to $\G$ and for any $k\le K$ we have
\[
V_k(g^\star)\ \le\ \Theta(w^\star)^2-1,\qquad\text{with equality when }w^\star_k>0 .
\]
Finally, we have
\[
\min_{g\in\G}\ \max_{k\le K}V_k(g)\ =\ \Theta(w^\star)^2-1 .
\]
\end{theorem}

Over all densities, the minimax problem is therefore a concave maximisation, whereas Buta and Doss \cite{BD11} judged the minimax design hopeless when the proposals are chosen among candidates for a family of Bayes factors. Every component with $w^\star_k>0$ attains the optimal coefficient of variation. It is \cref{thm:main} for the error criterion $\Phi(C)=\max_kC_{kk}/m_k^2$, with $M^\star=\mathrm{diag}(w^\star/m^2)$; the proof in \cref{app:proofs} is self-contained. For indicators it explains why mixtures of the conditional densities $f|_{A_k}$ work well. In the next proposition, $p_k:=\Prob_f(X\in A_k)>0$, the sets $A_k$ may overlap, and $g_\alpha:=\sum_k\alpha_k\,f|_{A_k}$ as before.

\begin{proposition}\label{prop:ind}
Under the assumptions of \cref{thm:minimax}, with $\psi_k=\ind_{A_k}$, for any sets $(A_k)$ the equal mixture satisfies
\[
\max_{k\le K}V_k(g_{\mathbf 1/K})\ \le\ K-1 .
\]
In addition, we have the following:
\begin{itemize}[leftmargin=*,itemsep=0pt]
\item When the sets are disjoint, $g_{\mathbf 1/K}$ is optimal and the optimal value is $K-1$.
\item When the sets are nested, $A_1\supset\dots\supset A_K$, the optimum is the mixture $g_{\alpha^\star}$ with
\[
\alpha_k^\star=\frac{p_k\,(s_k-s_{k-1})}{\Theta(w^\star)}\qquad\text{and}\qquad s_j:=\Big(\sum_{k\le j}\frac{w_k^\star}{p_k^2}\Big)^{1/2},\quad s_0:=0.
\]
\end{itemize}
\end{proposition}

Hence, as anticipated in \cref{ex:two}, for probabilities in relative error the optimum differs from a mixture of conditional densities only on overlaps of non-nested sets, where it uses $(\sum_kw_k/p_k^2)^{1/2}$ instead of $\sum_k\alpha_k/p_k$. The following example shows that this difference is negligible.

\begin{figure}[h]
\centering
\includegraphics[width=\textwidth]{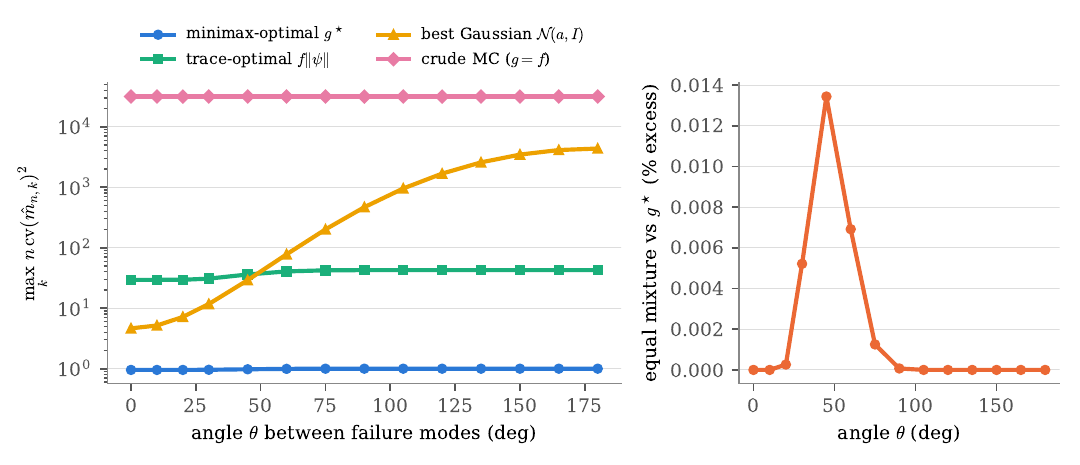}
\caption{Two failure modes at angle $\theta$. Left: $\max_kV_k=\max_kn\cv(\widehat m_{n,k})^2$ for several auxiliary densities (log scale). Right: excess of the equal mixture over the minimax optimum, in percent.}
\label{fig:sweep}
\end{figure}

\paragraph{Several failure modes.} We illustrate \cref{thm:minimax,prop:ind} on two failure modes. For $f=\mathcal N(0,I_d)$, consider the two half-spaces $A_k:=\{x\in\R^d:u_k^\top x\ge t_k\}$, $k=1,2$, with $t_1=3$, $t_2=4$ and unit vectors $u_1,u_2$ at angle $\theta$, so that $p_1=1.35\times10^{-3}$ and $p_2=3.17\times10^{-5}$; the sets are nested for $\theta=0^\circ$ and disjoint for $\theta=180^\circ$. Over $\theta\in[0^\circ,180^\circ]$ (\cref{fig:sweep}), the minimax optimum of \cref{thm:minimax} has $\max_kV_k$ between $0.954$ (nested) and $1$ (disjoint), as \cref{prop:ind} predicts. The equal mixture is within $0.014\%$ of it at every angle. The other densities are worse by factors from $5$ to $3\times10^4$ for this error criterion. The density $f\|\psi\|$, optimal for the trace error criterion, is $30$ to $43$ times worse, since it puts little mass on the rarer event. The best Gaussian $\mathcal N(a,I)$ with identity covariance, a classical IS density for a single event, is $5$ to $4\,400$ times worse, and crude Monte Carlo about $3\times10^4$ times worse. For four nested half-spaces $A_k=\{x(1)\ge k+1\}$, $k=1,\dots,4$ ($p$ from $2.3\times10^{-2}$ to $2.9\times10^{-7}$), the minimax optimum has $V_k=2.82$ for all $k$, against $2.96$ for the equal mixture, $12.9$ for the best $\mathcal N(a,I)$ and $4\times10^4$ for $f\|\psi\|$; the relative-error losses $\sum_k|x(k)/p_k|^r$ ($r=1,2,4$) and $\max_k|x(k)/p_k|$ all give densities within a few percent of each other.

For covariance matrices no such mixture is available, and both the loss and the family matter (\cref{sec:num}).

\subsection{A probability and its sensitivity}\label{sec:sens}

An expectation is often estimated together with its derivatives with respect to parameters of the law, as in the likelihood ratio method of gradient estimation \cite{Gly90} and in Bayesian sensitivity analysis \cite{BD11}. In reliability analysis, the sensitivities of a failure probability to the parameters of the input distribution and of the limit-state function are reported together with the probability \cite{PBS18}. When $f=f_\vartheta$ depends on a distribution parameter $\vartheta$, the classical score-function identity gives $\partial_\vartheta\Prob_\vartheta(X\in A)=\E_f\{\ind_A(X)\,\partial_\vartheta\log f_\vartheta(X)\}$, so that the pair (probability, sensitivity) is a vector of expectations with $\psi=\ind_A\,(1,\partial_\vartheta\log f_\vartheta)$. Its natural error criterion is relative, since the two components have different units, and \cref{thm:main} then gives the optimal density in closed form.

\begin{example}\label{ex:sens}
Let $f=\mathcal N(0,I_2)$, let $A:=\{x:x(1)\ge t+(c/2)\,x(2)^2\}$ be a parabolic failure set, i.e.\ $A=\{x:G(x)\le0\}$ for the limit-state function $G(x)=t+(c/2)\,x(2)^2-x(1)$, and let $\vartheta=\tau$ be the standard deviation of $X(2)$, at its nominal value $\tau=1$. Then $\partial_\tau\log f=x(2)^2-1$, and
\[
\psi(x)=\ind_A(x)\,\big(1,\ x(2)^2-1\big),\qquad m=\big(\Prob_f(X\in A),\ \partial_\tau\Prob_f(X\in A)\big).
\]
For the error criterion $\Phi(C):=(C_{11}/m_1^2+C_{22}/m_2^2)/2$, the mean squared coefficient of variation of the two estimates, which does not depend on the units of $\tau$, $\Phi$ is linear and \cref{thm:main} gives $M^\star=\mathrm{diag}(1/m_1^2,1/m_2^2)/2$ and
\[
g^\star\ \propto\ f\,\ind_A\,\Big[\frac1{m_1^2}+\frac{\{x(2)^2-1\}^2}{m_2^2}\Big]^{1/2}.
\]
The score changes sign on $A$, at $x(2)=\pm1$, and $g^\star$ keeps mass there, where the density $f|\psi_2|$, optimal for the sensitivity alone, vanishes: the latter gives the probability estimate an infinite variance. For $t=3$ and $c=1$, we have $m_1=6.41\times10^{-4}$ and $m_2=-5.01\times10^{-4}$, and \cref{tab:sens} compares auxiliary densities, with all integrals reduced to one dimension by integrating out $x(1)$.
\begin{table}[h]
\centering\small
\begin{tabular}{@{}lccc@{}}
\toprule
auxiliary density & $n\cv^2$ of the probability & $n\cv^2$ of the sensitivity & regret\\
\midrule
$f|_A$, i.e.\ exact failure samples & $0$ & $0.147$ & $1.29$\\
$f|\psi_2|$, optimal for the sensitivity alone & $\infty$ & $0.056$ & $\infty$\\
$f\|\psi\|$, trace optimum & $0.010$ & $0.106$ & $1.02$\\
$g^\star$ & $0.019$ & $0.095$ & $1$\\
\bottomrule
\end{tabular}
\caption{\Cref{ex:sens}, $t=3$ and $c=1$: squared relative errors and regrets of auxiliary densities.}
\label{tab:sens}
\end{table}

Here the regret of a density is its risk divided by that of $g^\star$, and $f|_A$ is the idealised density of exact failure samples with $\Prob_f(X\in A)$ known, as in \cref{sec:smooth}. Over $c\in\{0.5,1,2\}$ and $t\in\{3,4\}$, the regret of the failure samples ranges from $1.22$ to $1.60$, increasing with the curvature $c$: reusing them for the sensitivity costs a fifth to three fifths more model evaluations. The trace optimum is within $3\%$ in all these cases only because $|m_2/m_1|$ lies between $0.64$ and $0.90$; it is not invariant under a change of units of the parameter, and, for $t=3$ and $c=1$, replacing $\psi_2$ by $10\,\psi_2$ or $100\,\psi_2$ raises its regret to $2.6$ or $5.4$, whereas $g^\star$ is unchanged.
\end{example}

\section{No closed form: numerical computations}\label{sec:num}

For the covariance matrix $\Sigma$ of \cref{sec:smooth}, the optimal densities of the operator and nuclear norms have no closed form, and the risks of all densities must be computed; we do so numerically when the nominal density of \cref{sec:smooth} is $f|_A$, with mean $\mu_A$ and covariance $\Sigma_A$. This is the reliability instance: any scheme that fits a Gaussian auxiliary density to a failure domain must estimate $\mu_A$ and $\Sigma_A$, and the cross-entropy (CE) method \cite{Rub97,RK04,GPS19} estimates at each iteration the mean and covariance of $f$ restricted to an intermediate domain, by IS with self-normalised weights $f/g$ on the samples in this domain, the last domain being $A$ \cite{BSS25}. We call the estimation of $\Sigma_A$ at the last iteration the covariance step of CE and $\mathcal N(\mu_A,\Sigma_A)$ the CE target; taking $\mu_A$ and $\Prob_f(X\in A)$ known isolates this step. We measure what is lost with other densities: the densities optimal for another loss and the target $f|_A$ itself (\cref{sec:numcov}), and parametric families (\cref{sec:param}). Risks are compared as in \cref{sec:closed}.

\subsection{Covariance matrices: the loss matters in low dimension}\label{sec:numcov}

We first compare the optimal densities of different losses with each other and with $f|_A$. We estimate $\Sigma_A$ ($\mu_A$ known) for two failure sets, in dimensions $d=6$ ($K=21$) and $d=20$ ($K=210$):
\[
\begin{aligned}
&\text{half-space} && A=\{x(1)\ge2\}, && \Sigma_A=\mathrm{diag}(0.114,1,\dots,1),\\
&\text{paraboloid} && A=\{x(1)\ge2+\tfrac12\{x(2)^2+x(3)^2\}\}, && \Sigma_A=\mathrm{diag}(0.152,0.262,0.262,1,\dots,1).
\end{aligned}
\]
For the half-space, $\sigma_1=0.114$ is the variance of a standard normal variable conditioned to exceed $2$, and conditioning on $A$ does not affect the other coordinates. We call failure direction the direction $e_1$, along which the failure set extends to infinity. More generally, following \cite{BMS25}, we call influential the directions along which $f|_A$ differs from $f$ ($e_1$ for the half-space, $e_1,e_2,e_3$ for the paraboloid), and bulk directions the others, along which $f|_A$ coincides with $f$. We use the Frobenius, Stein/KL, Fr\'echet and precision losses of \cref{lem:local}, and the operator and nuclear norms of $\delta$, i.e.\ the largest absolute eigenvalue and the sum of the absolute eigenvalues, whose $M^\star_L$ is obtained from \eqref{eq:price}. For these two norms $\Phi_L$ is positively homogeneous of degree $1/2$, hence not convex, so that the densities reported as optimal are fixed points of \cref{thm:loss}, i.e.\ stationary points, which we could not certify as global minimisers; their regrets in \cref{tab:cov,tab:dim,tab:robust} are relative to these fixed points.

\begin{table}[h]
\centering\small
\begin{tabular}{@{}lcccccc@{}}
\toprule
optimal for $\downarrow$ / evaluated on $\to$ & Frobenius & Stein & Fr\'echet & precision & operator & nuclear\\
\midrule
Frobenius & \textbf{1} & 1.129 & 1.006 & 4.452 & 1.000 & 1.001\\
Stein/KL & 1.059 & \textbf{1} & 1.038 & 1.891 & 1.047 & 1.018\\
Fr\'echet & 1.004 & 1.060 & \textbf{1} & 3.273 & 1.006 & 0.999\\
precision & 1.782 & 1.444 & 1.714 & \textbf{1} & 1.413 & 1.292\\
operator norm & 1.001 & 1.168 & 1.012 & 5.091 & \textbf{1} & 1.003\\
nuclear norm & 1.000 & 1.104 & 1.003 & 4.015 & 1.002 & \textbf{1}\\
$g=f|_A$ &  1.479 & 1.500 & 1.465 & 4.602 & 1.270 & 1.201\\
\bottomrule
\end{tabular}
\caption{Half-space, $d=6$: cross-regrets, i.e.\ regret for the column loss of the density optimal for the row loss. Values below $1$ reflect Monte Carlo error ($\approx0.3\%$, with $2\times10^5$ draws, \cref{app:num}).}
\label{tab:cov}
\end{table}

In low dimension the losses fall into three groups (\cref{tab:cov}). The absolute-error losses (Frobenius, Fr\'echet, operator, nuclear), which do not rescale the errors by $\Sigma_A$, share one optimal density; the relative-error loss Stein/KL, which measures errors relative to $\Sigma_A$, differs from them by $2$ to $17\%$, depending on the direction of the comparison, and is the most robust choice; the precision loss, which weights the entry $(1,1)$ by $h_{11}=\sigma_1^{-4}$, needs a very different density (a regret of $1.9$ to $5.1$ for the densities of the other losses). Sampling from $f|_A$ itself, the zero-variance density for the single probability $\Prob_f(X\in A)$, is never optimal here, as \cref{thm:loss} predicts: no loss makes $\sqrt{\psi^\top M\psi}$ constant.

These differences, however, are a low-dimensional effect (\cref{tab:dim}). The paraboloid at $d=6$ behaves like the half-space, but at $d=20$ Stein and the absolute-error losses agree to $1.5\%$ in both geometries, precision to within a factor $1.1$ to $1.6$, and $f|_A$ comes within $5$--$12\%$ of optimal except for precision. The unit-variance bulk directions dominate $\sqrt{\psi^\top M\psi}$ and are weighted identically by every loss, so $g^\star$ approaches $f|_A$ as $d$ grows; only losses that strongly weight the influential directions keep their own optimum.

\begin{table}[h]
\centering\small
\begin{tabular}{@{}lcccc@{}}
\toprule
& \multicolumn{2}{c}{half-space} & \multicolumn{2}{c}{paraboloid}\\
\cmidrule(lr){2-3}\cmidrule(lr){4-5}
& $d=6$ & $d=20$ & $d=6$ & $d=20$\\
\midrule
Stein vs absolute-error & $1.02$--$1.17$ & $\le1.01$ & $1.03$--$1.24$ & $\le1.015$\\
precision vs others & $1.3$--$5.1$ & $1.07$--$1.56$ & $1.2$--$2.1$ & $1.06$--$1.26$\\
sampling from $f|_A$ & $1.2$--$4.6$ & $1.05$--$1.62$ & $1.2$--$2.1$ & $1.05$--$1.32$\\
\bottomrule
\end{tabular}
\caption{Dependence on the dimension and the geometry: worst cross-regrets between the non-parametric optima of the six losses, and regret of $f|_A$; ranges are over the losses.}
\label{tab:dim}
\end{table}

\subsection{Parametric families: the family matters in both dimensions tested}\label{sec:param}

An adaptive scheme does not sample the optimal densities of \cref{sec:numcov} directly: it works within a parametric family, Gaussian for CE. We now measure what this restriction costs. We consider Gaussians aligned with the symmetry of the problem,
\[
g=\mathcal N\big\{a\,e_1,\mathrm{diag}(\gamma_1,\gamma_{23},\gamma_{23},\gamma_\perp,\dots,\gamma_\perp)\big\},\qquad a\in\R,\ \gamma_1,\gamma_{23},\gamma_\perp>0,
\]
with $\gamma_{23}=\gamma_\perp$ for the half-space. The mean-shift family fixes $\gamma_1=\gamma_{23}=1$ and only adapts $a$ and $\gamma_\perp$.

A constraint on $\gamma_1$ follows from the tails. Here the nominal density is $f|_A$, and $C(g)$ is finite only if $(f|_A)^2\|\psi\|^2/g$ is integrable, so that along any direction in which $A$ is unbounded, the tails of $g$ must be heavier than those of $(f|_A)^2\|\psi\|^2$, up to an integrable factor; since $f$ is Gaussian, this gives an explicit bound. The failure set is unbounded along $e_1$, and on it the integrand of $S(g)$ behaves, up to polynomial factors, like $f^2/g\propto\exp\big[-\{1-1/(2\gamma_1)\}\,x(1)^2+O\{x(1)\}\big]$. Hence $C(g)$ is finite only if $\gamma_1\ge1/2$ (for $a>0$, the linear term ensures integrability at $\gamma_1=1/2$): the variance of a Gaussian auxiliary density must be at least half the variance of $f$ in any direction along which the failure set is unbounded. For the half-space, $C(g)$ is a product of one-dimensional integrals and is computed exactly (\cref{app:num}). The mixtures we fit have two Gaussian components of this form: a wide one, constrained to $\gamma_1\ge1/2$ so that the risk stays finite, and a narrow one, whose parameters are free; in all our fits the latter places itself at the boundary of $A$. The tail condition also holds with self-normalised weights, since centring does not change the growth of $\psi$.

\begin{table}[h]
\centering\small
\begin{tabular}{@{}lcccc@{}}
\toprule
& \multicolumn{2}{c}{half-space} & \multicolumn{2}{c}{paraboloid}\\
\cmidrule(lr){2-3}\cmidrule(lr){4-5}
& $d=6$ & $d=20$ & $d=6$ & $d=20$\\
\midrule
CE target $\mathcal N(\mu_A,\Sigma_A)$ & $\infty$ & $\infty$ & $\infty$ & $\infty$\\
mean shift only & $1.9$--$4.0$ & $1.8$--$3.4$ & $2.8$--$9.0$ & $2.7$--$7.7$\\
best Gaussian & $1.6$--$2.8$ & $1.6$--$2.4$ & $1.8$--$3.6$ & $1.7$--$3.0$\\
best two-component mixture & $1.2$--$1.8$ & $1.2$--$1.4$ & $1.5$--$2.6$ & $1.4$--$2.1$\\
$\gamma_1$ of the best Gaussian & $1/2$ & $1/2$ & $1/2$ & $1/2$\\
\bottomrule
\end{tabular}
\caption{Regret of parametric densities, i.e.\ their risk divided by that of $g^\star$, where $\infty$ means an infinite risk, and variance along $e_1$ of the best Gaussian; ranges are over the six losses.}
\label{tab:robust}
\end{table}

\begin{figure}[t]
\centering
\includegraphics[width=\textwidth]{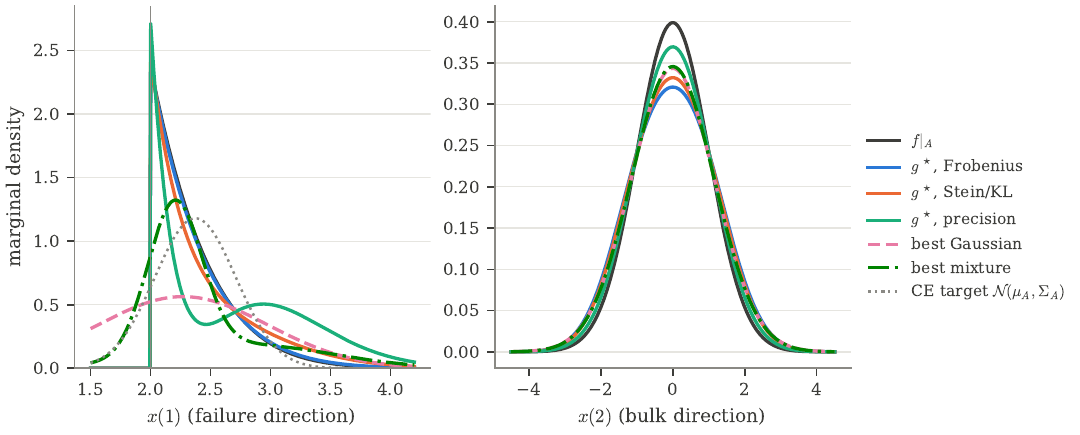}
\caption{Half-space, $d=6$: marginal densities along the failure direction $x(1)$ (left) and a bulk direction $x(2)$ (right). Optimal densities $g^\star$ for three losses, best Gaussian and best two-component mixture for the Stein loss, and the CE target. The vertical line is the boundary of $A$.}
\label{fig:dens}
\end{figure}

Three facts stand out (\cref{tab:robust}):
\begin{itemize}
\item The Gaussian target of CE has infinite risk, since its variance along the failure direction is $\sigma_1<1/2$; and every loss pins the best Gaussian at $\gamma_1=1/2$, the edge of finite variance. A safety margin is cheap: $\gamma_1=0.6$ costs $4$--$9\%$. Along $e_2$ and $e_3$, the directions in which the paraboloid curves and $f|_A$ has lighter-than-Gaussian tails, the Gaussian can concentrate ($\gamma_{23}\approx0.3$--$0.45$).
\item The Gaussian family multiplies the risk by $2.2$ to $3.6$ for the four quadratic losses, which is also the factor on the sample size, and by $1.55$ to $2.04$ for the operator and nuclear norms, that is, by $2.4$ to $4.2$ in sample size, in both dimensions tested, while the loss-dependence largely disappears within the family (for the half-space at $d=6$, the best Gaussians for all but the precision loss coincide to $0.5\%$). This factor comes from the profile of $f|_A$ along the failure direction, truncated and exponential-like, which no Gaussian matches (\cref{fig:dens}).
\item Two-component mixtures recover a third to three quarters of the gap to $g^\star$, more for the half-space than for the paraboloid, always with the same structure: a wide component at $\gamma_1=1/2$ with an inflated bulk, and a narrow component (variance $0.01$--$0.07$ along $e_1$) just inside the boundary of $A$.
\end{itemize}

\paragraph{Self-normalised weights.} CE weights its samples by $f/g$ normalised to sum to one, with $\mu_A$ and $\Prob_f(X\in A)$ unknown. The resulting estimator of $\Sigma_A$ has asymptotic covariance $C(g)$ computed for the centred function $\psi-m$ (\cref{sec:setting}), so that its optimal density is $f|_A\{(\psi-m)^\top M(\psi-m)\}^{1/2}$ up to normalisation. For the half-space with the Stein and Frobenius losses, in dimensions $6$ and $20$, the conclusions above carry over: relative to the optimal risk, estimated by Monte Carlo from $4\times10^5$ draws of $f|_A$, the density optimal with known $\Prob_f(X\in A)$ loses at most $4\%$ under self-normalised weights, $f|_A$ loses $10$ to $39\%$, and the best Gaussian multiplies the risk by $2.2$ to $2.4$, again with $\gamma_1=1/2$, a margin $\gamma_1=0.6$ costing $7$ to $9\%$. These Gaussian risks are computed exactly, as in \cref{app:num}.

\Cref{fig:dens} shows the mechanism for the half-space at $d=6$. Along the failure direction, every optimal density jumps at the boundary of $A$ and decays like $f|_A$; the precision loss weights $f|_A$ essentially by $y(1)^2/\sigma_1^2$, which is large far from $\mu_A(1)$, and adds a second mode in the tail. A Gaussian with $\gamma_1=1/2$ can only approximate the jump by spreading mass on both sides of the boundary, and its mass outside $A$ is wasted. The narrow component of the mixture sits at the boundary and reproduces the jump, while the wide component covers the tail. In the bulk directions all optimal densities are slightly wider than $f$, which is the bulk inflation of the best Gaussians. The CE target $\mathcal N(\mu_A,\Sigma_A)$ is too narrow along the failure direction and puts too little mass both at the boundary and in the tail.

\FloatBarrier
\section{Discussion}\label{sec:discussion}

\Cref{thm:main,thm:loss} reduce the choice of one auxiliary density for several expectations to the choice of one $K\times K$ matrix: a subgradient of the error criterion, or half the averaged Hessian of the loss. This matrix is explicit for smooth losses and nearly explicit for partitions, and in general it can be learned from a pilot sample.

\paragraph{When the error criterion matters.} The answer depends on the scales of the components. When they range over orders of magnitude, as for the probabilities of rare cells, the error criterion changes the optimal density by orders of magnitude, and the sum of the variances, which favours the largest components, is far from optimal for relative errors (\cref{prop:alloc,tab:hist}). For several events in relative error, the equal mixture of their conditional densities is essentially optimal (\cref{thm:minimax,prop:ind}). For covariance matrices, the loss matters in low dimension and for the precision loss, and its influence fades with the dimension (\cref{sec:numcov}).

\paragraph{Consequences for adaptive schemes.} An adaptive scheme that fits a parametric density to $f|_A$ estimates a vector of expectations at each iteration, and our results apply to this step. For the covariance step of CE, the Gaussian target $\mathcal N(\mu_A,\Sigma_A)$ has infinite risk: along the directions in which the failure set is unbounded, for Gaussian $f$, the variance of a Gaussian auxiliary density must be at least half the variance of $f$ (\cref{sec:param}). Within the Gaussian family the loss matters little, and mixtures with a component at the boundary of $A$ recover a third to three quarters of the gap to $g^\star$.

\paragraph{Open problems.} When $\Phi_L$ is not convex, as for norms, \cref{thm:loss} only gives a necessary condition, and conditions that exclude spurious stationary points are missing. Sampling $g_M$ is the bottleneck in practice (\cref{rem:pilot}), and parametric families or samplers adapted to $g_M$, embedded in an adaptive scheme, remain to be designed. Finally, for the self-normalised estimator the centring $m$ is itself estimated, and its finite-sample analysis is left for future work.

\appendix
\section{Numerical details}\label{app:num}

This appendix gives the computational details and the Monte Carlo checks behind \cref{sec:closed,sec:num}. We write $\varphi_{a,\Gamma}$ for the density of $\mathcal N(a,\Gamma)$ and $\FN$ for the standard normal c.d.f.

\paragraph{Failure probabilities and histograms.} For $b\in\{0,1\}^K$ we denote by $A_b$ the atom of the sets $(A_k)$ on which $\ind_{A_k}=b(k)$ for any $k\le K$. Every candidate of the form $g=f\,u(\ind_{A_1},\dots,\ind_{A_K})$ with $u:\{0,1\}^K\to[0,\infty)$ has
\[
V_k(g)=p_k^{-2}\sum_{b:\,b(k)=1}\frac{\Prob_f(X\in A_b)}{u(b)}-1,
\]
so all risks reduce to the $2^K$ atom probabilities $\Prob_f(X\in A_b)$, computed with $\FN$ in one coordinate and a trapezoid rule in the other (marginals exact to $3\times10^{-15}$). The dual of \cref{thm:minimax} was solved by exponentiated gradient and matched a direct primal solve over all $u\ge0$ to relative error $10^{-8}$. Sampling $g^\star$ exactly by rejection from the $\sqrt{w^\star}$-weighted mixture ($\theta=45^\circ$, $n=2000$, $5000$ replications) gave $n\cv(\widehat m_{n,k})^2=0.955$ against the exact $0.977$ (s.e.\ $0.02$); dual weights estimated from a pilot of $500$ draws were within $6\times10^{-4}$ of $w^\star$.

\paragraph{Pilot sample (\cref{rem:pilot}).}\label{app:pilot} The model is $y=u^\top x$ in dimension $d=10$, with $u$ a fixed random unit vector, treated as a black box, and the cells are $(-\infty,0)$, $[j,j+1)$ for $0\le j<J$ and $[J,\infty)$, with $J=5$ ($K=7$) or $J=7$ ($K=9$, $\Prob_f(Y\ge7)\approx1.3\times10^{-12}$). Multilevel mean-shift CE for $\{y\ge J\}$ (family $\mathcal N(\mu,I)$, $1000$ draws per level, quantile $0.1$; full-covariance CE degenerated at this sample size, and the improved CE of \cite{PGS19} was not tried) gives the levels $g_1,\dots,g_T$; the pilot density is the equal mixture of $f,g_1,\dots,g_T$ with $n_0=2000$, which put at least $129$ draws in every cell over all replications. The final density is $0.05f+0.95\sum_k\alpha_k\mathcal N(\widehat\mu_k,\widehat\Sigma_k)$, with $(\widehat\mu_k,\widehat\Sigma_k)$ the IS moments of $f|_{A_k}$ from the pilot sample and $\alpha$ from the fixed point, or $\alpha_k=1/K$ (equal mixture). The CE density is $0.05f+0.95g_T$ with $n_0+n$ draws, and crude Monte Carlo uses the whole budget. \Cref{tab:pilot} reports risks over $1000$ replications divided by the exact risk of crude Monte Carlo at the same total budget; the empirical risk of crude Monte Carlo is not used because it is dominated by rare hits of the tail cells.
\begin{table}[h]
\centering\small
\begin{tabular}{@{}lcccc@{}}
\toprule
& \multicolumn{2}{c}{$K=7$} & \multicolumn{2}{c}{$K=9$}\\
\cmidrule(lr){2-3}\cmidrule(lr){4-5}
total budget (final $n$) & $7000$ ($1000$) & $11000$ ($5000$) & $12000$ ($5000$) & $27000$ ($20000$)\\
\midrule
\multicolumn{5}{@{}l}{$\chi^2$ loss}\\
CE density & $23.8$ & $16.9$ & $18.5$ & $13.8$\\
equal mixture & $8.68$ & $2.78$ & $3.02$ & $1.65$\\
fixed point & $4.63$ & $1.45$ & $1.36$ & $0.74$\\
exact $g_M$, same overhead & $2.74$ & $0.90$ & $0.74$ & $0.41$\\
\midrule
\multicolumn{5}{@{}l}{total variation loss}\\
CE density & $7.67$ & $6.44$ & $7.01$ & $6.00$\\
equal mixture & $4.76$ & $2.71$ & $3.25$ & $2.39$\\
fixed point & $3.27$ & $1.87$ & $2.01$ & $1.47$\\
exact $g_M$, same overhead & $2.50$ & $1.46$ & $1.51$ & $1.13$\\
\bottomrule
\end{tabular}
\caption{Finite-sample risks with a pilot sample, divided by the exact risk of crude Monte Carlo at the same total budget (below $1$: better than crude Monte Carlo). Standard errors are at most $3\%$ of the values.}
\label{tab:pilot}
\end{table}
The asymptotic regrets of crude Monte Carlo are $1.03$ (total variation) and $2.48$ to $3.31$ ($\chi^2$), which bounds what any density can gain; the regret of the estimated $\alpha$, with exact sampling of the cells, was at most $1.0015$, and the estimated dual bound exceeded the exact minimum in half of the replications.

\paragraph{Probability and sensitivity (\cref{ex:sens}).} The optimal value matches the dual bound $\Lambda(M^\star)$ to $10^{-12}$ in relative terms.

\paragraph{Covariance matrices.} Draws of $f|_A$ are exact: by inverse c.d.f.\ for the half-space and, for the paraboloid, rejection on $(x(2),x(3))$ followed by a truncated normal for $x(1)$, with $\Prob_f(X\in A)$, $\mu_A$ and $\Sigma_A$ from one-dimensional quadrature. With nominal density $f|_A$ and $Q:=\psi^\top M\psi$, we have $C(g_M)=Z(M)\,\E_{f|_A}(\psi\psi^\top/\sqrt Q)-mm^\top$ with $Z(M)=\E_{f|_A}(\sqrt Q)$, both estimated from $2\times10^5$ draws. For the half-space $A=\{x(1)\ge2\}$ and Gaussian $g$, the entry of $S(g)$ indexed by the pairs $(i,j)$ and $(i',j')$ is, up to the $\sqrt2$ factors, $\prod_{l\le d}T_l(\nu_l)$, with
\[
T_1(\nu):=\int_2^\infty\frac{\{\varphi_{0,1}(x)/P_A\}^2}{\varphi_{a,\gamma_1}(x)}\,\{x-\mu_A(1)\}^\nu\,\mathrm{d}x\qquad\text{and}\qquad T_l(\nu):=\frac{\sqrt{\gamma_\perp}\,\E(\zeta^\nu)}{(2-1/\gamma_\perp)^{(\nu+1)/2}}\quad(l\ge2),
\]
where $P_A:=\Prob_f(X\in A)$, $\zeta\sim\mathcal N(0,1)$ and $\nu_l$ is the number of occurrences of $l$ among $i,j,i',j'$; this agreed with Monte Carlo ($2\times10^6$ draws) to $0.5$--$1.7\%$. Mixture risks for the half-space at $d=6$ are estimated from $4\times10^5$ exact draws of $f|_A$; all other Gaussian and mixture risks from $2\times10^5$ draws, with parameters fitted on an independent sample of $10^5$ draws; for the half-space at $d=20$ they reproduce the exact risks of the best Gaussians to $0.3\%$. These estimates are consistent, but they have infinite variance when the variance along $e_1$ of $g$, or of the wide component of a mixture, is below $2/3$, since $(f|_A)^3\|\psi\|^4/g^2$ is then not integrable along $e_1$. This is the case for the best Gaussians on the paraboloid, at $\gamma_1=1/2$, and possibly for the mixtures, so that no standard error is reported for them; the agreement with the exact risks above and the re-evaluations below serve instead. Mixture optima were re-evaluated on independent samples and changed by less than $0.3\%$; relaxing the lower bound on the narrow component's variance changed the risk by at most $2\%$.

\section{Supporting results and proofs}\label{app:proofs}\label{sec:wellposed}

This appendix states the supporting results used in the main text, each just before its proof, and gathers the proofs of all the results, in logical order, so that each proof uses only results proved before it.

\begin{lemma}\label{lem:convex}
For any $g_0,g_1\in\G$ and $t\in[0,1]$, the density $(1-t)g_0+tg_1$ belongs to $\G$ and we have
\[
S\big\{(1-t)g_0+tg_1\big\}\ \preceq\ (1-t)\,S(g_0)+t\,S(g_1).
\]
In addition, under condition \cnd{C}{A2}, if $\Phi$ is convex then $g\mapsto\Phi\{C(g)\}$ is convex on $\G$.
\end{lemma}

\begin{proof}[Proof of \cref{lem:convex}]
For any $v\in\R^K$ we have $v^\top S(g)v=\int f^2(v^\top\psi)^2/g$, and $u\mapsto a/u$ is convex on $(0,\infty)$ for any $a\ge0$. The second assertion follows from the monotonicity and the convexity of $\Phi$.
\end{proof}

Consider the following strengthening of condition \cnd{C}{A2}:
\begin{itemize}[leftmargin=3.4em,labelindent=0pt,labelwidth=2.8em,labelsep=0.6em,align=left,itemsep=2pt]
\item[\cndef{Cinf}{A4}] Condition \cnd{C}{A2} is met, $\Phi$ is lower semicontinuous, and $\Phi(C)\to\infty$ as $\|C\|\to\infty$ with $C\in\Sp$.
\end{itemize}
Condition \cnd{Cinf}{A4} is met by the four error criteria listed after condition \cnd{Cplus}{A3} as soon as $W\in\mathcal S_K$ and $m_k\neq0$ for any $k$.

\begin{proposition}\label{prop:exist}
Assume \cnd{I}{A1} and \cnd{Cinf}{A4}, and that $f\|\psi\|$ is not almost everywhere zero. Then $g\mapsto\Phi\{C(g)\}$ attains its minimum on $\G$.
\end{proposition}

\begin{proof}[Proof of \cref{prop:exist}]
The density $g_0:=f\|\psi\|/\int f\|\psi\|$ belongs to $\G$, since $S(g_0)$ is the product of $\int f\|\psi\|$ and $\int_{\{\psi\neq0\}}f\,\psi\psi^\top/\|\psi\|$, and $\|\psi\psi^\top\|/\|\psi\|=\|\psi\|$. Moreover $\Phi\{C(g)\}\ge\Phi(0)>-\infty$ for any $g\in\G$ by \cnd{C}{A2}. Let $(g_n)_{n\ge1}$ be a minimising sequence. By \cnd{Cinf}{A4} the matrices $S(g_n)$ are bounded, so that, extracting a subsequence, $S(g_n)\to\bar S$ and, by Prokhorov's theorem on the one-point compactification of $\R^d$, the probability measures $g_n(x)\,\mathrm{d}x$ converge narrowly to some $\nu$. We write $\nu=g(x)\,\mathrm{d}x+\nu_s+\nu(\{\infty\})\,\delta_\infty$, with $\nu_s$ carried by a Lebesgue-null Borel set $N\subset\R^d$.

Fix $v\in\R^K$ and set $b:=f\,v^\top\psi\in L^1$. For any continuous and compactly supported $\omega$ on $\R^d$ we have $2b\omega-\omega^2g_n\le b^2/g_n$ pointwise, hence $2\int b\omega-\int\omega^2g_n\le v^\top S(g_n)v$, and in the limit
\begin{equation}\label{eq:dualS}
2\int b\,\omega-\int\omega^2\,g-\int\omega^2\,\mathrm{d}\nu_s\ \le\ v^\top\bar Sv .
\end{equation}
For $T>0$, let $\omega_T:=\mathrm{sign}(b)\,(T\wedge|b|/g)$, with $|b|/g:=\infty$ on $\{g=0\}$; one checks that $2b\omega_T-\omega_T^2g\ge(b^2/g)\,\ind_{\{|b|\le Tg\}}+T|b|\,\ind_{\{|b|>Tg\}}\ge0$. Let $\varepsilon>0$. By the outer regularity of the Lebesgue measure there is an open set $U\supset N$ such that $\int_U(|b|+g)\le\varepsilon$, and by Lusin's theorem, applied on $\R^d$ to $\omega_T\,\ind_{B\setminus U}$, for a large ball $B$, and the finite measure $(|b|+g)\,\mathrm{d}x+\nu_s$, there is a continuous compactly supported $\omega$ with $|\omega|\le T$ that coincides with $\omega_T\,\ind_{B\setminus U}$ outside a set $E$ with $\int_E(|b|+g)+\nu_s(E)\le\varepsilon$ and $\int_{\R^d\setminus B}(|b|+g)\le\varepsilon$ for the ball $B$. Then $\int\omega^2\,\mathrm{d}\nu_s\le T^2\varepsilon$, and \eqref{eq:dualS} yields
\[
\int_{\{|b|\le Tg\}}\frac{b^2}{g}+T\int_{\{|b|>Tg\}}|b|\ \le\ v^\top\bar Sv+c\,(T+T^2)\,\varepsilon
\]
for some universal constant $c$. Letting $\varepsilon\to0$, the second term on the left shows that $\int_{\{g=0\}}|b|\le v^\top\bar Sv/T$ for any $T>0$, hence $b=0$ almost everywhere on $\{g=0\}$; letting then $T\to\infty$ in the first term, we obtain $\int b^2/g\le v^\top\bar Sv$. In particular $g>0$ almost everywhere on $\{f\|\psi\|>0\}$, and $t:=\int g>0$ since $f\|\psi\|$ is not almost everywhere zero.

The density $g^\star:=g/t$ thus satisfies $S(g^\star)=t\,S(g)\preceq\bar S$, so that $g^\star\in\G$ and $C(g^\star)\preceq\bar S-mm^\top=\lim_nC(g_n)$. By the monotonicity and the lower semicontinuity of $\Phi$ we conclude that
\[
\Phi\{C(g^\star)\}\ \le\ \Phi\big\{\lim_nC(g_n)\big\}\ \le\ \liminf_n\Phi\{C(g_n)\}\ =\ \inf_{g\in\G}\Phi\{C(g)\} .
\]
This ends the proof of the proposition.
\end{proof}

\begin{proof}[Proof of \cref{thm:main}]
By the Fenchel--Young inequality, $\Phi\{C(g)\}\ge\tr\{MC(g)\}-\Phi^*(M)$, and by \eqref{eq:cs} we have $\tr\{MC(g)\}=\tr\{MS(g)\}-m^\top Mm\ge Z(M)^2-m^\top Mm$, with equality if and only if $g=g_M$ when $Z(M)>0$; if $Z(M)=0$, then $\tr\{MS(g)\}=0$ for any $g$ and the bound is trivial. This proves \eqref{eq:weak}. The function $M\mapsto Z(M)^2=\inf_{g\in\G}\tr\{MS(g)\}$ is an infimum of linear functions; the infimum over $\G$ equals $Z(M)^2$ even when $g_M\notin\G$, since for any $g_0\in\G$ and $\varepsilon\in(0,1)$ the density $g_\varepsilon:=(1-\varepsilon)g_M+\varepsilon g_0$ belongs to $\G$ and, by convexity of $g\mapsto f^2\psi^\top M\psi/g$, satisfies $\tr\{MS(g_\varepsilon)\}\le(1-\varepsilon)Z(M)^2+\varepsilon\tr\{MS(g_0)\}$. Hence $Z^2$ is concave and, $\Phi^*$ being convex, so is $\Lambda$. If $M^\star$ is self-consistent and $g^\star:=g_{M^\star}$, then the Fenchel--Young inequality is an equality at $(C(g^\star),M^\star)$, since $M^\star\in\partial\Phi\{C(g^\star)\}$, and so is \eqref{eq:cs}, since $g=g^\star$. Hence $\Phi\{C(g^\star)\}=\Lambda(M^\star)$, and \eqref{eq:weak} shows that $g^\star$ is a minimiser and $M^\star$ a maximiser of $\Lambda$.
\end{proof}

\begin{proof}[Proof of \cref{prop:nec}]
Set $C^\star:=C(g^\star)$. Since $\Phi$ is nondecreasing, $\Phi(C^\star+t\Delta)\ge\Phi(C^\star)$ for any $\Delta\in\Sp$ and $t>0$, so that $\tr(M\Delta)\ge0$ and $M\in\Sp$. For any $g_1\in\G$ and $t\in(0,1]$, \cref{lem:convex} and the monotonicity of $\Phi$ yield
\begin{align*}
\Phi(C^\star)\ &\le\ \Phi\big[C\{(1-t)g^\star+tg_1\}\big]\ \le\ \Phi\big[C^\star+t\,\{S(g_1)-S(g^\star)\}\big]\\
&=\ \Phi(C^\star)+t\,\tr\big[M\{S(g_1)-S(g^\star)\}\big]+o(t),
\end{align*}
where the expansion is licit because $C^\star+t\,\{S(g_1)-S(g^\star)\}=(1-t)\,C^\star+t\,C(g_1)$ belongs to $\Sp$. This implies that $\tr\{MS(g^\star)\}\le\tr\{MS(g_1)\}$ for any $g_1\in\G$. On the other hand, $g_M$ itself may fail to belong to $\G$, but for any $\varepsilon\in(0,1)$ the density $g_\varepsilon:=(1-\varepsilon)g_M+\varepsilon g^\star$ satisfies $S(g_\varepsilon)\preceq S(g^\star)/\varepsilon$, so that $g_\varepsilon\in\G$, and by convexity of $u\mapsto f^2\psi^\top M\psi/u$, the integrand being $0$ on $\{g_M=0\}$, we have
\[
\tr\big\{MS(g_\varepsilon)\big\}\ \le\ (1-\varepsilon)\,Z(M)^2+\varepsilon\,\tr\big\{MS(g^\star)\big\}\ \longrightarrow\ Z(M)^2\quad\text{as }\varepsilon\to0 .
\]
Since $Z(M)>0$, we conclude that $\tr\{MS(g^\star)\}=Z(M)^2$, which is the equality case of \eqref{eq:cs}.
\end{proof}

Consider the following condition on the loss $L$ and an auxiliary density $g$:
\begin{itemize}[leftmargin=3.4em,labelindent=0pt,labelwidth=2.8em,labelsep=0.6em,align=left,itemsep=2pt]
\item[\cndef{U}{B2}] For a given $g\in\G$, the sequence of random variables $n^{\kappa/2}L(\widehat m_n-m)$, $n\ge1$, is uniformly integrable when $X_1,X_2,\dots$ are i.i.d.\ from $g$.
\end{itemize}
Condition \cnd{U}{B2} is the usual requirement to pass from convergence in law to convergence of risks; in case (a) it is met as soon as $\kappa<2$, since $\E\{\|\sqrt n(\widehat m_n-m)\|^2\}=\tr C(g)$, or as soon as $\ell\psi$ has a moment of order $\kappa+\varepsilon$ under $g$ for some $\varepsilon>0$, by the Marcinkiewicz--Zygmund inequality.

\begin{lemma}\label{lem:risk}
Assume \cnd{I}{A1}, \cnd{L}{B1} and \cnd{U}{B2} for some $g\in\G$. Then we have
\[
\lim_{n\to\infty}\ n^{\kappa/2}\ \E\{L(\widehat m_n-m)\}\ =\ \Phi_L\big\{C(g)\big\}.
\]
\end{lemma}

\begin{proof}[Proof of \cref{lem:risk}]
By the central limit theorem, $\sqrt n(\widehat m_n-m)$ converges in law to $C(g)^{1/2}\xi$. In case (a), $n^{\kappa/2}L(\widehat m_n-m)=L\{\sqrt n(\widehat m_n-m)\}$ and we conclude with the continuous mapping theorem. In case (b), $L(x)=L_0(x)+o(\|x\|^2)$ as $x\to0$, so that $nL(\widehat m_n-m)-L_0\{\sqrt n(\widehat m_n-m)\}$ tends to $0$ in probability, and we conclude in the same way. In both cases condition \cnd{U}{B2} turns convergence in law into convergence of the expectations.
\end{proof}

\begin{lemma}[\cite{And55,Pri58,Ste81}]\label{lem:price}
Assume \cnd{L}{B1}. Then $\Phi_L$, extended by $+\infty$ outside $\Sp$, satisfies condition \cnd{C}{A2}. In addition, $\Phi_L$ is differentiable on $\mathcal S_K$ with, for any $C\in\mathcal S_K$,
the gradient \eqref{eq:price}. Moreover, $\Phi_L$ is differentiable relative to $\Sp$ at any $C\in\Sp$, with gradient $H/2$, in case (b); and, if $L_0(x)=\sum_jc_j|v_j^\top x|^r$ with $c_j>0$, $v_j\in\R^K\setminus\{0\}$ and $r\ge1$, at any $C\in\Sp$ such that $v_j^\top Cv_j>0$ for any $j$, with gradient $\sum_j(r/2)\,c_j\,\E(|\zeta|^r)\,(v_j^\top Cv_j)^{(r-2)/2}\,v_jv_j^\top$, where $\zeta\sim\mathcal N(0,1)$. Finally, $\Phi_L$ is continuous on $\Sp$, and condition \cnd{Cinf}{A4} is met as soon as $L(x)>0$ for any $x\neq0$ in case (a), or $H\in\mathcal S_K$ in case (b).
\end{lemma}

\begin{proof}[Proof of \cref{lem:price}]
Monotonicity is Anderson's inequality \cite{And55}: for $C\preceq C'$, write $\mathcal N(0,C')$ as the law of $X+Y$ with $Y\sim\mathcal N(0,C'-C)$ independent of $X$; since $L_0$ is even and quasi-convex we have $\E\{L_0(X+y)\}\ge\E\{L_0(X)\}$ for any $y$. For \eqref{eq:price} (Price's theorem \cite{Pri58}), the Gaussian density $\varphi_C:=\varphi_{0,C}$ satisfies $\partial\varphi_C/\partial C=\nabla^2\varphi_C/2$; we integrate by parts, which is licit since $L_0$ has polynomial growth, and we use $\nabla^2\varphi_C=\varphi_C\,(C^{-1}xx^\top C^{-1}-C^{-1})$ \cite{Ste81}. The gradient of a differentiable nondecreasing function belongs to $\Sp$. In case (b), $\Phi_L(C)=\tr(HC)/2$ is linear. For $L_0(x)=\sum_jc_j|v_j^\top x|^r$, the variable $v_j^\top C^{1/2}\xi$ is $\mathcal N(0,v_j^\top Cv_j)$, so that $\Phi_L(C)=\E(|\zeta|^r)\sum_jc_j(v_j^\top Cv_j)^{r/2}$ on $\Sp$; the right-hand side is differentiable on the open set $\{C\in\Ms:v_j^\top Cv_j>0\text{ for any }j\}$ with the stated gradient, which gives differentiability relative to $\Sp$. Continuity on $\Sp$ follows from the continuity of $C\mapsto C^{1/2}$ and dominated convergence. In case (a), $L(x)\ge c_-\|x\|^\kappa$ with $c_-:=\min_{\|u\|=1}L(u)>0$, so that $\Phi_L(C)\ge c_-\,\E(|\zeta|^\kappa)\,\|C\|^{\kappa/2}$ with $\zeta\sim\mathcal N(0,1)$, by projecting $C^{1/2}\xi$ on a top eigenvector of $C$; in case (b), $\Phi_L(C)=\tr(HC)/2\ge\lambda_{\min}(H)\|C\|/2$.
\end{proof}

\begin{proof}[Proof of \cref{thm:loss}]
By \cref{lem:price}, $\Phi_L$ satisfies \cnd{C}{A2}, so that the first assertion is \cref{prop:nec}, and the expression of the gradient on $\mathcal S_K$ is \eqref{eq:price}. When $\Phi_L$ is convex, it satisfies \cnd{Cplus}{A3} by \cref{lem:price}, and a gradient $M$ relative to $\Sp$ at $C$ is a subgradient: by convexity $\Phi_L(C')\ge\Phi_L(C)+\tr\{M(C'-C)\}$ for any $C'\in\Sp$, and $\Phi_L=+\infty$ outside $\Sp$. The second assertion is then \cref{thm:main}.
\end{proof}

\begin{proposition}\label{prop:ui}
Assume \cnd{I}{A1}, assume that $f\|\psi\|$ is not almost everywhere zero, and let $M\in\mathcal S_K$. Then $g_M\in\G$ and we have
\[
\|\ell\,\psi\|\ \le\ Z(M)\,\lambda_{\min}(M)^{-1/2}\qquad g_M\text{-almost everywhere}.
\]
Consequently, under condition \cnd{L}{B1}, condition \cnd{U}{B2} is met for $g=g_M$, and \cref{lem:risk} yields
\[
\lim_{n\to\infty}n^{\kappa/2}\,\E\{L(\widehat m_n-m)\}\ =\ \Phi_L\big\{C(g_M)\big\}.
\]
\end{proposition}

\begin{proof}[Proof of \cref{prop:ui}]
On $\{g_M>0\}$ we have\[\ell\psi=\frac{Z(M)\,\psi}{\sqrt{\psi^\top M\psi}}\qquad\text{and}\qquad\psi^\top M\psi\ge\lambda_{\min}(M)\|\psi\|^2,\]which yields the bound. It implies that\[\|S(g_M)\|\ \le\ \int\|\ell\psi\|^2g_M\ \le\ Z(M)\,\lambda_{\min}(M)^{-1/2}\int f\|\psi\|\ <\ \infty,\] and $g_M>0$ exactly on $\{f\|\psi\|>0\}$, so that $g_M\in\G$. Since the summands $\ell(X_i)\psi(X_i)-m$ are bounded and centred, the Marcinkiewicz--Zygmund inequality yields $\sup_n\E\{\|\sqrt n(\widehat m_n-m)\|^s\}<\infty$ for any $s\ge2$. In case (a) of \cnd{L}{B1} we have $n^{\kappa/2}L(\widehat m_n-m)\le c_+\,\|\sqrt n(\widehat m_n-m)\|^\kappa$ with $c_+:=\max_{\|u\|=1}L(u)$. In case (b), the Taylor expansion at $0$ and the growth condition yield $L(x)\le c'(\|x\|^2+\|x\|^{q})$ with $q:=c_0\vee2$, so that $nL(\widehat m_n-m)\le c'\big\{\|\sqrt n(\widehat m_n-m)\|^2+\|\sqrt n(\widehat m_n-m)\|^q\big\}$, since $n^{1-q/2}\le1$. In both cases the sequence is bounded in $L^2$, hence uniformly integrable.
\end{proof}

\begin{proof}[Proof of \cref{lem:local}]
Write $\Sigma=\mathrm{diag}(\sigma)$ and $B:=\Sigma^{-1/2}\delta\,\Sigma^{-1/2}$, so that $\tr B^2=\sum_{i,j}\delta_{ij}^2/(\sigma_i\sigma_j)$. For the Stein loss, $\tr B-\log\det(I+B)=(\tr B^2)/2+O(\|B\|^3)$. The Kullback--Leibler divergence between Gaussian laws with the same mean is half the Stein loss. The Bhattacharyya distance is\[\tfrac12\log\det(I+B/2)-\tfrac14\log\det(I+B)=\tfrac1{16}\tr B^2+O(\|B\|^3).\] For the precision loss, $\widehat\Sigma^{-1}-\Sigma^{-1}=-\Sigma^{-1}\delta\Sigma^{-1}+O(\|\delta\|^2)$, whose entries are $-\delta_{ij}/(\sigma_i\sigma_j)$. For the log-Euclidean loss, the Daleckii--Krein formula \cite[Theorem V.3.3]{Bha97} gives $\{\mathrm{D}\log(\Sigma)[\delta]\}_{ij}=\delta_{ij}\,(\log\sigma_i-\log\sigma_j)/(\sigma_i-\sigma_j)$. For the Fr\'echet loss, $W_2^2=\tr\widehat\Sigma+\tr\Sigma-2\tr R$ with $R:=(\Sigma^{1/2}\widehat\Sigma\Sigma^{1/2})^{1/2}$. Along $\widehat\Sigma=\Sigma+t\delta$ we have $R(0)=\Sigma$ and, differentiating $R^2=\Sigma^2+t\,\Sigma^{1/2}\delta\Sigma^{1/2}$ twice,
\[
\dot R_{ij}=\frac{\sqrt{\sigma_i\sigma_j}\,\delta_{ij}}{\sigma_i+\sigma_j},\qquad \tr\ddot R=-\sum_i\frac{(\dot R^2)_{ii}}{\sigma_i}=-\sum_{i,j}\frac{\sigma_j\,\delta_{ij}^2}{(\sigma_i+\sigma_j)^2}=-\frac12\sum_{i,j}\frac{\delta_{ij}^2}{\sigma_i+\sigma_j}.
\]
Since $\tr\dot R=(\tr\delta)/2$, the first-order terms cancel and $W_2^2=-t^2\tr\ddot R+o(t^2)$. Finally, with $x_{ii}=\delta_{ii}$ and $x_{ij}=\sqrt2\,\delta_{ij}$, the local form reads $x^\top Hx/2$ with $H=2\,\mathrm{diag}\{(h_{ij})_{i\le j}\}$, so that $M^\star_L=H/2$ by \cref{lem:price}, and $\psi^\top M^\star_L\psi=\sum_ih_{ii}y(i)^4+2\sum_{i<j}h_{ij}y(i)^2y(j)^2$. This ends the proof of the lemma.
\end{proof}

\begin{proof}[Proof of \cref{prop:alloc}]
Let $\zeta\sim\mathcal N(0,1)$. The loss is separable, and differentiating $\Phi_L(C)=\E(|\zeta|^r)\sum_kC_{kk}^{r/2}p_k^{-\beta}$ shows that $M^\star_L$ is diagonal with\[(M^\star_L)_{kk}=\tfrac r2\,\E(|\zeta|^r)\,C_{kk}^{(r-2)/2}p_k^{-\beta}.\] On $A_k$ we have $\psi^\top M^\star_L\psi=(M^\star_L)_{kk}$, so that $g^\star=g_\alpha$ with $\alpha_k\propto p_k\{(M^\star_L)_{kk}\}^{1/2}$. Substituting $C_{kk}=p_k^2(1-\alpha_k)/\alpha_k$ yields \eqref{eq:alloc-exact}.
\end{proof}

\begin{proof}[Proof of \cref{thm:minimax}]
The function $w\mapsto\sqrt{Q_w(x)}$ is the square root of a nonnegative linear function, which yields the concavity of $\Theta$. Write $Q:=Q_{w^\star}$, $\Theta^\star:=\Theta(w^\star)$ and $R_k:=\psi_k^2/m_k^2$.

\emph{Support.} If $B:=\{Q=0,\ \psi_j\neq0\}$ had positive $f$-measure, then $w_j^\star=0$, and $w_\varepsilon:=(1-\varepsilon)w^\star+\varepsilon e_j$ would satisfy $\Theta(w_\varepsilon)\ge\sqrt{1-\varepsilon}\,\Theta^\star+\sqrt\varepsilon\int_Bf|\psi_j|/|m_j|>\Theta^\star$ for small $\varepsilon$, a contradiction. So $g^\star>0$ on $\{f\|\psi\|>0\}$.

\emph{Equalisation.} By concavity of $\sqrt\cdot$ and monotone convergence, the directional derivative
\[
\Theta'(w^\star;e_k-w^\star)=\tfrac12\int fR_k/\sqrt Q-\tfrac12\Theta^\star
\]
is nonpositive, and it vanishes when $w_k^\star>0$, since the opposite direction is then feasible and $t\mapsto\Theta\{w^\star+t(e_k-w^\star)\}$ is differentiable at $0$: the integrand is differentiable on $\{Q>0\}$, $R_k=0$ $f$-almost everywhere on $\{Q=0\}$ by the support step, and monotone convergence applies on both sides. Since $V_k(g^\star)=\Theta^\star\int fR_k/\sqrt Q-1$, this yields $V_k(g^\star)\le\Theta^{\star2}-1$, with equality when $w_k^\star>0$. In particular the diagonal of $S(g^\star)$ is finite, hence $S(g^\star)$ is finite by the Cauchy--Schwarz inequality, and $g^\star\in\G$.

\emph{Lower bound.} For any $g\in\G$, by the Cauchy--Schwarz inequality we have
\[
\max_{k\le K}V_k(g)\ \ge\ \sum_{k\le K}w_k^\star V_k(g)\ =\ \int\frac{f^2Q}{g}-1\ \ge\ \Theta^{\star2}-1 .
\]
This ends the proof of the theorem.
\end{proof}

\begin{proof}[Proof of \cref{prop:ind}]
On $A_k$ we have $g_{\mathbf 1/K}\ge f/(Kp_k)$, so that $\int\ind_{A_k}f^2/g_{\mathbf 1/K}\le Kp_k^2$, which yields the first bound. For disjoint sets $\Theta(w)=\sum_k\sqrt{w_k}$, which is maximised at $\mathbf 1/K$ with $\Theta(\mathbf 1/K)^2-1=K-1$. For nested sets, on $A_j\setminus A_{j+1}$ we have $\sqrt{Q_{w^\star}}=s_j$, which is nondecreasing in $j$; we conclude by writing $s_j=\sum_{k\le j}(s_k-s_{k-1})$ and $\ind_{A_k}=\sum_{j\ge k}\ind_{A_j\setminus A_{j+1}}$.
\end{proof}

\end{document}